\pdfoutput=1
\documentclass[11pt, reqno]{article}
\usepackage{jheppub}
\usepackage{epsfig}
\usepackage{amssymb}
\usepackage{amsmath}
\usepackage{mathrsfs}
\usepackage{hyperref}
\usepackage{multirow}
\usepackage{feynmp-auto}
\usepackage[utf8]{inputenc}

\def\be{\begin{equation}}
\def\ee{\end{equation}}
\def\ba{\begin{eqnarray}}
\def\ea{\end{eqnarray}}
\newcommand{\bea}{\begin{eqnarray}}
\newcommand{\eea}{\end{eqnarray}}

\def\eps{\epsilon}

\newcommand{\fwboxL}[2]{\text{\makebox[#1][l]{$#2$}}}

\newcommand{\floor}[1]{\lfloor #1 \rfloor}

\newcommand{\lsim}{\mathrel{\hbox{\rlap{\lower.55ex \hbox{$\sim$}} \kern-.3em \raise.4ex \hbox{$<$}}}}

\newcommand{\Gauss}{ {}_2F_1 }
\newcommand{\ol}{\Omega^{(L)}}
\newcommand{\li}{\text{Li}}
\newcommand{\ub}[2]{ \underbrace{\scriptstyle #1,\ldots,#1}_{#2} }
\renewcommand{\a}{\alpha}

\title{A Novel Algorithm for Nested Summation and Hypergeometric Expansions}

\author{Andrew~J.~McLeod$^1$,}
\author{Henrik~Jessen~Munch$^{1,2}$,}
\author{Georgios~Papathanasiou$^3$,}
\author{and Matt~von~Hippel$^1$}

\affiliation{$^1$ Niels Bohr International Academy, Blegdamsvej 17,
2100 Copenhagen, Denmark}

\affiliation{$^2$  II. Institut für Theoretische Physik,  Universität Hamburg, Luruper Chaussee 149, 22761 Hamburg}

\affiliation{$^3$ DESY Theory Group, DESY Hamburg, Notkestra{\ss}e 85,
D-22607 Hamburg, Germany}

\preprint{DESY 20-079}

\abstract{We consider a class of sums over products of $Z$-sums whose arguments differ by a symbolic integer. Such sums appear, for instance, in the expansion of Gauss hypergeometric functions around integer indices that depend on a symbolic parameter. We present a telescopic algorithm for efficiently converting these sums into generalized polylogarithms, $Z$-sums, and cyclotomic harmonic sums for generic values of this parameter. This algorithm is illustrated by computing the double pentaladder integrals through ten loops, and a family of massive self-energy diagrams through $\mathcal{O}(\epsilon^{6})$ in dimensional regularization. We also outline the general telescopic strategy of this algorithm, which we anticipate can be applied to other classes of sums.}

\emailAdd{amcleod@nbi.ku.dk}\emailAdd{henrikjessenmunch@gmail.com}\emailAdd{georgios.papathanasiou@desy.de}\emailAdd{mvonhippel@nbi.ku.dk}

\begin{document}
\hypersetup{pageanchor=false}

\maketitle
\hypersetup{pageanchor=true}
\begin{fmffile}{feyndiags}

\section{Introduction}

Hypergeometric functions appear ubiquitously in physics---including as they do special cases such as Legendre polynomials and Bessel functions---and in particular are known to appear in dimensionally-regularized perturbative quantum field theory. This ubiquity is due in part to the extensive class of second-order differential equations that hypergeometric functions solve, which has been the subject of much dedicated research. While many aspects of these functions are well understood as a result of this research, they can still prove unwieldy in practical calculations.

Generalized (or multiple) polylogarithms also appear in many places in quantum field theory. While they are less general than hypergeometric functions, they are correspondingly under better control; in particular, a great deal of progress has been made leveraging motivic aspects of these functions, considered as iterated integrals on the moduli space of the Riemann sphere with marked points~\cite{Goncharov:2001iea,Brown:2009qja,Brown1102.1312,Brown:2015fyf}. For instance, all functional relations between generalized polylogarithms can in principle be exploited with the use of the coaction~\cite{Goncharov:2005sla,2011arXiv1101.4497D}, as applied in~\cite{Brown:2011ik,Duhr:2011zq,Duhr:2012fh,Bourjaily:2019igt}. There also exist public codes for the efficient numerical evaluation of polylogarithms, for instance in {\sc GiNaC}~\cite{Bauer:2000cp, Vollinga:2004sn}.

In some cases, the hypergeometric functions that appear in physics can be expressed as infinite sums of polylogarithms, allowing us to leverage this polylogarithmic technology. This occurs, for instance, in the case of one-loop dimensionally regulated Feynman integrals~\cite{vanNeerven:1985xr,Bern:1993kr,Anastasiou:1999ui}. For finite values of the dimensional regularization parameter $\epsilon$, these integrals can be expressed in terms of hypergeometric functions, while their series expansion around small values of $\epsilon$ can be written in terms of polylogarithms. While the sum representation of hypergeometric functions is a useful starting point for performing this expansion, explicitly converting the expansion coefficients into polylogarithms can prove to be a nontrivial task. Great progress in this respect has been made in~\cite{Moch:2001zr}, where several summation algorithms, covering a large number of cases, were presented (see also~\cite{Weinzierl:2002hv,Moch2006759,Huber:2005yg} for their computer algebra implementation, as well as~\cite{Kalmykov:2006hu} for related work). Even so, these algorithms are not always capable of carrying out the series expansion around \emph{symbolic} integer indices of hypergeometric functions. In the context of dimensionally-regularized Feynman integrals these symbolic integers essentially correspond to propagators raised to generic powers.

In this paper, we extend the work of~\cite{Moch:2001zr} by presenting an algorithm for explicitly evaluating series expansions of Gauss hypergeometric functions taking the form
\begin{equation}
	\label{eq:2F1_constraint}
	\Gauss(k + \eps_1, l +  \eps_2, m + \eps_3|x), 
	\qquad|\epsilon_i| \ll 1, \quad \text{with}\,\, k-m\,\,\text{and}\,\,l\,\,\text{fixed integers}\, ,
\end{equation}
in terms of well-known classes of functions. Importantly, since we place only two conditions on the integer parts of the hypergeometric indices, \eqref{eq:2F1_constraint} is a function of an \emph{arbitrary symbolic integer}, which we will generally denote $\alpha$, on top of the complex argument $x$. Hence, our expansion may prove especially useful when hypergeometric functions of the form~\eqref{eq:2F1_constraint} appear as part of a larger expression in which the symbolic integer $\alpha$ is summed over, or simply because this expansion can be evaluated a single time and then used for different values of $\alpha$.\footnote{It is also well known that all $\Gauss$ functions whose arguments differ by integer shifts may be expressed in terms of a basis of two functions in this family. In this sense, our algorithm could be thought of as the explicit expansion not only of the basis, but also of the \emph{coefficients} of any other function in the family. Also note that software such as {\sc {Hyperdire}}~\cite{Bytev:2011ks} can carry out this reduction to a basis only when $\alpha$ is a fixed integer, which will not be the case here.}

More generally, we show how sums of the form
\begin{equation} \label{eq:general_class_sums}
	\sum_{n=1}^{N} \frac{x^n}{n^p(n+\a)^q} Z_{m_1,\ldots,m_d}(n{-}1|y_1,\ldots,y_{d}) Z_{r_1,\ldots,r_h}(n{+}\a{-}1|z_1,\ldots,z_{h}) \, ,
\end{equation}
can be efficiently evaluated for symbolic values of $\{x, y_1, \dots, y_d, z_1, \dots, z_h\} \in \mathbb{C}$ and $\{N,\a\} \in \mathbb{Z}_{\geq 0}$, where (as we will review in section $\ref{sec:Zsums}$) $Z$-sums are given by
\be \label{eq:ZsumIntro}
	Z_{m_1, \ldots, m_d}(N | x_1, \ldots, x_d)= \sum_{N \ge i_1 > i_2 > \dots > i_d > 0} \frac{x_1^{i_1}}{i_1^{m_1}} \cdots \frac{x_d^{i_d}}{i_d^{m_d}}\, .
\ee
In particular, the sums in~\eqref{eq:general_class_sums} evaluate to generalized polylogarithms and $Z$-sums when $N$ is infinite, and cyclotomic harmonic sums~\cite{Ablinger:2011te} for generic $N$. The sums appearing in the expansion \eqref{eq:2F1_constraint} correspond to the special case in which $x_i=y_i=m_i=1$ and $N \to \infty$.

We achieve this result with the help of \emph{telescoping} identities. In their simplest guise, these take the form
\begin{align}
	\label{SimpleTele}
	\Phi(\a) = \sum_{\mu=1}^{\a-1} \Delta \Phi(\mu) + \Phi(1)\, , \qquad \quad	\Delta \Phi(\mu) = \Phi(\mu{+}1) - \Phi(\mu)\, ,
\end{align}
and can be used to compute $\Phi(\a)$ if the quantity $\Delta \Phi(\alpha)$ is simpler than the original sum. We will make use of a generalized identity of the form~\eqref{SimpleTele}, in which the analog of $\Delta \Phi(\alpha)$ has lower depth---that is, involves fewer nested sums---than the original sum.  This will allow us to leverage a recursion in the depth of sums taking the form \eqref{eq:general_class_sums}, and in this manner bring them to a form that may be evaluated with the existing algorithms of~\cite{Moch:2001zr}.


There exist of course well-established (creative) telescoping methods for deriving recurrence relations of symbolic sums, see for example \cite{10.2307/67570,10.1016/S0747-7171(08)80044-2,article}. While not necessarily restricted to this case, these recurrences are typically with respect to the upper summation variable (or variables linearly related to the latter), i.e.~with respect to $N$ in equation~\eqref{eq:general_class_sums}. The novelty of our approach is that we not only consider $\alpha$ in the latter formula to be a symbolic integer, but that we also derive a recursion with respect to the latter. In particular, at the heart of our algorithm lies the master formula \eqref{eq:Spq_recursion}, which to the best of our knowledge has never been considered before, and is not a simple application of existing methods.

That being said, the algorithm of~\cite{Anzai:2012xw} (see also~\cite{Blumlein:2010zv} for earlier related work) in principle offers another alternative route for converting the sums in~\eqref{eq:general_class_sums} into nested sums for generic values of $\a$. Nevertheless, we find that this algorithm proves overly computationally expensive in the cases we consider, due to the fact that it generates spurious term-wise divergences  at intermediate steps in the calculation. These divergences require regularization, and cancel out in the final result. For sums of the form~\eqref{eq:general_class_sums}, we therefore view our approach as simpler and more efficient.

In order to illustrate the utility of our algorithm, we apply it to two examples. First, we consider the class of double pentaladder integrals introduced in \cite{ArkaniHamed:2010kv}. A compact generating function for these integrals was derived in~\cite{Caron-Huot:2018dsv}, which gives rise to a sum representation involving products of $\Gauss$ functions that fall into the class~\eqref{eq:2F1_constraint}. Using our algorithm, we explicitly evaluate these integrals in terms of generalized polylogarithms through ten loops. Second, we consider dimensionally-regularized one-loop self-energy diagrams for generic masses and propagator powers. In the limit of zero external momentum, these integrals become expressible in terms of $\Gauss$ functions~\cite{Anastasiou:1999ui}, and our algorithm can be used to simultaneously expand families of integrals that have different propagator powers around $4 - 2 \eps$ dimensions. We carry out the expansion and resummation of one such family of self-energy diagrams through $\mathcal{O}(\epsilon^6).$

This paper is organized as follows. In section \ref{sec:Zsums}, we begin by reviewing the aspects of $Z$-sums and generalized polylogarithms that will be relevant for our analysis. We then initiate the series expansion of Gauss hypergeometric functions taking the form~\eqref{eq:2F1_constraint}, and deduce the types of nested sums they give rise to. In section \ref{sec:Algorithm}, we present the general strategy of our algorithm, and state the telescoping recursion it gives rise to for sums of the form~\eqref{eq:general_class_sums}. Finally, we derive this recursion in a simplified setting, for compactness of presentation. Sections \ref{sec:double_pentaladder} and \ref{sec:self_energy} deal with the applications of our algorithm to the double pentaladder integrals and the massive self-energy integrals, respectively. In section \ref{sec:conclusions} we conclude. We also include two appendices, in which we derive the most general form of our recursion and present an example involving cyclotomic harmonic sums.

We accompany the arXiv pre-print of this paper with illustrative {\sc Mathematica} code which evaluates sums of the form~\eqref{eq:general_class_sums} into generalized polylogarithms, $Z$-sums, and cyclotomic harmonic sums. This consists of a package \texttt{Telescoping.wl} and a notebook of examples \texttt{Examples.nb}. The code employs the publicly available package {\sc HarmonicSums}~\cite{ablinger2010computer,Ablinger:2013hcp,Ablinger:2018cja,Ablinger:2016lzr,Ablinger:2014rba,Ablinger:2013eba,Ablinger:2015gdg,Ablinger:2018pwq,Ablinger:2011te,Ablinger:2019mkx,Ablinger:2013cf,Ablinger:2014bra,Blumlein:2009ta,Remiddi:1999ew,Vermaseren:1998uu} for 
standard operations on harmonic sums, in particular stuffle products, sum synchronization, $S$- to $Z$-sum conversion and differentiation, which we review in the next section, as well as numerics. We also include computer-readable files containing expressions for the double pentaladder integrals through six loops.

\section{\texorpdfstring{$Z$-sums, Polylogarithms, and $\Gauss$ Functions}{Z-sums, Polylogarithms, and 2F1 Functions}}\label{sec:Zsums}

We begin by reviewing the types of sums that arise when $\Gauss$ functions are expanded around integer values of their indices. Although the coefficients of these expansions are known to be expressible in terms of generalized polylogarithms around fixed integer values~\cite{Kalmykov:2007pf}, they can evaluate to the (more general) class of $Z$-sums~\cite{Moch:2001zr} around generic symbolic integers. We review this class of sums, and then describe how they relate to generalized polylogarithms and $\Gauss$ functions.

\subsection{\texorpdfstring{$Z$}{Z}-sums}\label{subsec:Zsums}
 
We give just a brief review of $Z$-sums, introducing notation and recalling the properties that will prove useful in later sections. The reader is referred to~\cite{Moch:2001zr} for more details. Starting from any integer $N$ and the initial definition
\begin{align} \label{eq:Z_sum_def_initial}
	Z(N) \equiv 
	\begin{cases} 
		1, & N \geq 0 \\
		0, & N < 0 \, ,
	\end{cases}
\end{align}
$Z$-sums are defined recursively in terms of pairs of variables $m_i \in \mathbb{Z}_+$ and $x_i \in \mathbb{C}$ by
\begin{align} \label{eq:Z_sum_def}
	Z_{m_1, \ldots, m_d}(N | x_1, \ldots, x_d) &\equiv
	\sum_{i=1}^{N} \frac{x_1^{i}}{i^{m_1}} Z_{m_2, \ldots, m_d}(i{-}1 | x_2, \ldots, x_d) \\
	&= \sum_{N \ge i_1 > i_2 > \dots > i_d > 0} \frac{x_1^{i_1}}{i_1^{m_1}} \cdots \frac{x_d^{i_d}}{i_d^{m_d}}\, . \label{eq:Z_sum_compact}
\end{align}
The depth of each $Z$-sum is defined to be its number of summation indices $d$, and its weight is defined to be $w = \sum_{i=1}^d m_i$. In general, we will adopt an abbreviated notation in which bold characters indicate multi-indices, for instance 
\begin{align}\label{eq:ZsumAbbrev}
Z_{\mathbf{m}}(N|\mathbf{x}) \equiv Z_{m_1,\dots,m_d}(N|x_1,\dots,x_d) \, ,
\end{align}
which leaves the depth of each sum implicit. We will also adopt the notation that the first entry of primed multi-indices has been dropped, for instance $\mathbf{m'} = m_2,\dots,m_{d}$, and denote the number of indices in $\mathbf{m}$ by $|\mathbf{m}|$.

The $Z$-sums obey a stuffle algebra as a consequence of the ability to split up unordered sums into nested ones. More precisely, a product of $Z$-sums that have the same upper summation limit $N$ (but which don't necessarily have the same depth) may be expressed in terms of $Z$-sums of higher depth by recursively applying
\begin{align}
	Z_{\mathbf{m}}(N|\mathbf{x}) \times Z_{\mathbf{n}}(N|\mathbf{y}) &= 
	\sum_{i=1}^{N} \frac{x_1^i}{i^{m_1}}Z_{\mathbf{m'}}(i{-}1|\mathbf{x'}) Z_{\mathbf{n}}(i{-}1|\mathbf{y}) \\ 
	&\hspace{1cm} + \sum_{i=1}^{N} \frac{{y_1}^i}{i^{n_1}}Z_{\mathbf{m}}(i{-}1|\mathbf{x}) Z_{\mathbf{n'}}(i{-}1|\mathbf{y'}) \nonumber \\ 
	&\hspace{1cm} + \sum_{i=1}^{N} \frac{(x_1 y_1)^i}{i^{m_1+n_1}}Z_{\mathbf{m'}}(i{-}1|\mathbf{x'}) Z_{\mathbf{n'}}(i{-}1|\mathbf{y'}) \,  \nonumber 
\end{align}
until the depth of the original $Z$-sums has been reduced to zero (after which new sums are built up using~\eqref{eq:Z_sum_def}). For example, we can reexpress
\begin{align}
	Z_{1,1}(N|x_1,x_2) \times Z_2(N|y) 	& =
	Z_{1,1,2}(N|x_1,x_2,y) + Z_{1,2,1}(N|x_1,y,x_2) + Z_{1,3}(N|x_1,x_2y) \nonumber \\ & \qquad +
	Z_{2,1,1}(N|y,x_1,x_2) + Z_{3,1}(N|x_1y,x_2) \, .
\end{align}
It is also interesting to note that this stuffle algebra has an associated coalgebra, and that together these structures form a Hopf algebra; however, the associated coproduct is not the coproduct usually encountered in Feynman integral calculations, which is associated with their mixed Tate Hodge structure~\cite{Brown:2015fyf} (see for instance~\cite{Weinzierl:2015nda}).

Identities also exist between sums with different summation bounds. For instance, relations between $Z$-sums with different upper bounds follow directly from~\eqref{eq:Z_sum_def}, namely 
\begin{align}
	\label{ZIdentity}
	Z_{\mathbf{m}}(N {+} M |\mathbf{x}) =
	Z_{\mathbf{m}}(N |\mathbf{x}) + \sum_{n=1}^{M} \frac{x_1^{N+n}}{(N+n)^{m_1}}Z_{\mathbf{m'}}(N {+} n {-} 1 |\mathbf{x'})
\end{align}
for $M,N \in \mathbb{Z}_+$ and $|\mathbf{m}|>0$. Equation~\eqref{ZIdentity} is particularly useful for \emph{synchronizing} a product of two $Z$-sums with different upper summation bounds, as it can be applied (iteratively) to replace one of the $Z$-sums with a linear combination of (products of) $Z$-sums with shifted summation bounds.

Additionally, the sums one encounters often have different lower summation bounds than allowed in~\eqref{eq:Z_sum_def}; in particular, we will see below that expansions of gamma functions that appear in the denominator of hypergeometric functions more naturally give rise to $S$-sums, which are closely related to $Z$-sums~\cite{Moch:2001zr}. Specifically, these sums satisfy
\begin{align}
	S_{\mathbf{m}}(N | \mathbf{x}) =
	\sum_{N \ge i_1 \ge i_2 \ge \dots \ge i_d \ge 1} \frac{x_1^{i_1}}{i_1^{m_1}} \cdots \frac{x_d^{i_d}}{i_d^{m_d}} \, ,
\end{align}
which can be compared to~\eqref{eq:Z_sum_compact}. The $S$-sums have similar algebraic properties to the $Z$-sums, but we prefer to work in terms of $Z$-sums as they are more directly related to generalized polylogarithms. $S$-sums can be converted to $Z$-sums iteratively using
\begin{align} \label{eq:S_to_Z_conversion}
	S_{\mathbf{m}}(N | \mathbf{x}) &= S_{m_1+m_2,\mathbf{m}'}(N| x_1 x_2, \mathbf{x}')  + \sum_{i_1 =1}^N \frac{x_1^{i_1}}{i_1^{m_1}} \sum_{i_2 =1}^{i_1-1} \frac{x_2^{i_2}}{i_2^{m_2}} S_{\mathbf{m}''}(i_2 | \mathbf{x}'') \, ,
\end{align}
which separates out the contribution to these sums in which pairs of summation indices are equal; this leaves summation bounds that fit the definition of $Z$-sums, at the cost of introducing a sum of lower depth (making it clear that this recursion terminates). Notice that we have used double primes to indicate dropping the first two indices in a multi-indix (for example $\mathbf{x}'' = x_3,\dots,x_d$). 

An important sub-class of the $Z$-sums are the Euler-Zagier sums~\cite{Zagier:1994}, which occur when all $x_i = 1$; similarly, $S$-sums with all arguments evaluated at unity reduce to the harmonic sums~\cite{Vermaseren:1998uu}. We abbreviate both cases using
\begin{align}
	Z_{\mathbf{m}}(N) &\equiv Z_{\mathbf{m}}(N|1, \ldots, 1)\, , \\
	S_{\mathbf{m}}(N) &\equiv S_{\mathbf{m}}(N|1, \ldots, 1)\, .
\end{align}
These sums appear naturally in the expansion of the gamma function and its reciprocal, and therefore also in the expansion of hypergeometric functions.

\subsection{Generalized Polylogarithms}

Many of the quantities encountered in quantum field theory can be expressed entirely in terms of generalized polylogarithms. This corresponds to expressing these quantities in terms of $Z$-sums in which $N \to \infty$, as
\begin{align} \label{eq:Z_sums_to_polylogs}
	\lim_{N \to \infty} Z_{m_1,\ldots,m_d}(N|x_1,\ldots,x_d) =
	\li_{m_d,\ldots,m_1}(x_d,\ldots,x_1)
\end{align}
reproduces the normal definition of generalized polylogarithms (note the reversal of indices and arguments)~\cite{Chen,G91b,Goncharov:1998kja,Remiddi:1999ew,Borwein:1999js,Moch:2001zr}. The depth and (transcendental) weight of generalized polylogarithms coincide with the definitions they were given above as $Z$-sums.

Generalized polylogarithms form a closed subalgebra within the $Z$-sums, and thereby inherit the algebraic properties of the larger space. Moreover, they can be given an integral definition
\begin{align} \label{eq:polylog_integral_def}
G_{x_1,\dots,x_n}(z) \equiv \int_0^z \frac{dt}{t-x_1} G_{x_2,\dots,x_n} \, , \qquad G_{\fwboxL{27pt}{{\underbrace{0,\dots,0}_{n}}}}(z) \equiv \frac{1}{n!} \log^n z \, ,
\end{align}
where $z, x_i \in \mathbb{C}$, and where the second definition accounts for the cases in which the first $n$ indices are zero (as the general definition diverges in these cases). This definition is related to the sum definition by
\begin{align}
\li_{m_d,\ldots,m_1}(x_d,\ldots,x_1) = (-1)^d \hspace{.06cm} G_{\fwboxL{28pt}{{\underbrace{0,\dots,0}_{m_1-1}}},\frac{1}{x_1},\hspace{.06cm}\text{\dots},\hspace{.06cm}\fwboxL{28pt}{{\underbrace{0,\dots,0}_{m_d-1}}},\frac{1}{x_d \cdots x_1}}(1) \, ,
\end{align} 
and allows polylogarithms to be analytically continued outside of the region of convergence $|x_i| < 1$. This representation also gives rise to a new set of identities analogous to the stuffle identities, corresponding to the ability to triangulate unordered integration ranges (coming from  products of polylogarithms) into sums over iterated integrals; these go by the name of shuffle identities. We refer the interested reader to the review~\cite{Duhr:2014woa} for further details.

In some of our examples, we will also make use of the harmonic polylogarithms (HPLs)~\cite{Remiddi:1999ew}. These are functions of a single argument $z$, and correspond to restricting $x_i \in \{0,1,-1\}$ in~\eqref{eq:polylog_integral_def}. The standard notation is given by
\begin{align}\label{eq:HPL}
H_{x_1,\dots,x_d}(z) \equiv (-1)^p G_{x_1,\dots,x_d}(z) \, ,
\end{align}
where $p$ is the number of indices $x_i$ that are 1. It can be useful to express functions in terms of HPLs when possible, due to the existence of several dedicated packages for their analytic and numerical evaluation~\cite{Gehrmann:2001pz,Maitre:2005uu,Maitre:2007kp,Buehler:2011ev}.

\subsection{Gauss Hypergeometric Function Expansions}\label{subsec:GaussExp}

Having introduced some of the necessary machinery in the previous subsections, let us now proceed with the expansion of the hypergeometric function \eqref{eq:2F1_constraint} mentioned in the introduction. Our starting point will be the series definition of the Gauss hypergeometric function,
\begin{equation}
	\label{eq:2F1Expansion}
	\Gauss(a,b,c|x) \equiv
	1 + \frac{\Gamma(c)}{\Gamma(a) \Gamma(b)} \sum_{n=1}^{\infty} 
	\frac{\Gamma(n+a) \Gamma(n+b)}{\Gamma(n+c)} \frac{x^n}{\Gamma(n+1)}  \, .
\end{equation}
This sum converges for $|x| < 1$, as can be seen by taking the $n\to\infty$ limit of consecutive terms in the series.

We will be interested in expanding the indices of Gauss hypergeometric functions around
\be\label{eq:abc_integer_expand}
a= k + \eps_1\,,\quad b=l +  \eps_2\,, \quad c=m + \eps_3\,,\quad \epsilon_i\to 0\,,
\ee
for integer $k$, $l$, and $m$. To this end, we make use of the identity
\begin{equation} \label{eq:gamma_eps_expansion}
	\Gamma(k+\eps) = \Gamma(1+\eps) \Gamma(k) \sum_{i=0}^{\infty} \eps^i Z_{\fwboxL{28pt}{{\underbrace{1,\dots,1}_{i}}}}(k-1) \, ,\quad k \in \mathbb{Z}_+\,.
\end{equation}
When we need to expand gamma functions in the denominator, we also employ
\begin{equation} \label{eq:gamma_eps_expansion_inverse}
	\bigg(\sum_{i=0}^{\infty} \eps^i Z_{\fwboxL{28pt}{{\underbrace{1,\dots,1}_{i}}}}(k-1)  \bigg)^{-1} = 
	\sum_{i=0}^{\infty} (-1)^i \eps^i S_{\fwboxL{28pt}{{\underbrace{1,\dots,1}_{i}}}}(k-1) \,,
\end{equation}
so as to place all nested sums in the numerator. With the help of these replacements, it is easy to see that the factor outside of the sum over $n$ in \eqref{eq:2F1Expansion} readily evaluates to $Z$- and $S$-sums in the limit~\eqref{eq:abc_integer_expand} (note also that the factors of $\Gamma(1+\eps)$ will cancel out of the overall expression). Therefore, the only nontrivial terms requiring evaluation in \eqref{eq:2F1Expansion} take the form
\be\label{eq:GeneralGaussSum}
\sum_{n=1}^{\infty} x^n \frac{\Gamma(n+k)\Gamma(n+l)}{\Gamma(n+m)\Gamma(n+1)}  Z_{\fwboxL{28pt}{{\underbrace{1,\dots,1}_{i_1}}}}(n+k-1) 
	Z_{\fwboxL{28pt}{{\underbrace{1,\dots,1}_{i_2}}}}(n+l-1)
	S_{\fwboxL{28pt}{{\underbrace{1,\dots,1}_{i_3}}}}(n+m-1)  \, 
\ee
at order $\epsilon_1^{i_1}\epsilon_2^{i_2}(-\epsilon_3)^{i_3}$ in the expansion.

Ideally, we would like to be able to evaluate sums of the type \eqref{eq:GeneralGaussSum} for general, symbolic values of all three integers $k$, $l$, and $m$. However, the presence of gamma functions in the above formula places significant obstacles on the telescoping methods mentioned in the introduction. For this reason, in what follows we specialize to a one-dimensional subspace of this 3D lattice, such that the aforementioned gamma functions reduce to polynomials in $n$. That is, we impose the constraint that
\begin{equation}\label{eq:GaussIndexRestriction}
k-m\,\,\text{and}\,\,l\,\text{are fixed integers}\,,
\end{equation}
which in turn allows us to replace
\begin{equation} \label{eq:gamma_function_ratio}
\frac{\Gamma(n+k)}{\Gamma(n+m)} = (n+m)(n+m+1) \cdots (n+k-m-1) \, ,
\end{equation}
 (where if $k>m$ we simply exchange $k$ and $m$) and similarly for $k\to l$, $m\to 1$.\footnote{As the $\Gauss$ function is symmetric under the exchange of its first two indices, we can equivalently impose this condition after swapping $l \leftrightarrow m$.} If this replacement results in a polynomial in $n$ in the numerator, we may use the differential operator introduced in~\cite{Moch:2001zr},
\begin{equation} \label{eq:derivative_operator}
	\mathbf{x^-} \equiv x \frac{\text{d}}{\text{d}x}\,,
\end{equation}
in order to reexpress each monomial in $n$ as
\begin{equation} \label{eq:derivative_operator_replacement}
	\sum_n x^n n^p =(\mathbf{x^-})^{p+1} \sum_n \frac{x^n}{n} \, .
\end{equation}
With this replacement, we can first evaluate the sum on the right-hand side, and then differentiate the result to evaluate the original sum. The differentiation of nested sums has already been implemented in software such as {\sc HarmonicSums}~\cite{ablinger2010computer,Ablinger:2013hcp,Ablinger:2018cja,Ablinger:2016lzr,Ablinger:2014rba,Ablinger:2013eba,Ablinger:2015gdg,Ablinger:2018pwq,Ablinger:2011te,Ablinger:2019mkx,Ablinger:2013cf,Ablinger:2014bra,Blumlein:2009ta,Remiddi:1999ew,Vermaseren:1998uu}.

After transforming the $S$-sums in \eqref{eq:GeneralGaussSum} to $Z$-sums with the help of~\eqref{eq:S_to_Z_conversion}, applying~\eqref{ZIdentity} to synchronize them, making the replacements~\eqref{eq:gamma_function_ratio}, and finally partial fractioning the denominator with respect to $n$, we find that only a single nontrivial class of sums needs to be evaluated: 
\begin{equation}
	\label{eq:SpecialGaussSum}
\sum_{n=1}^{\infty}\frac{x^n}{n^p(n+\a)^q} Z_{1,\text{\,\dots}, 1}(n-1) Z_{r_1,\text{\,\dots}, r_h}(n+\a-1)\, ,
\end{equation}
where $\alpha$ is the symbolic integer that remains after imposing the constraint~\eqref{eq:GaussIndexRestriction}. One may readily check that these sums are a special case of equation~\eqref{eq:ZsumIntro} for $x_i=y_i=m_i=1$, and $N \to \infty$.  In the next section, we present an algorithm for evaluating the more general class of sums, and therefore also for evaluating the sums~\eqref{eq:SpecialGaussSum} that appear in the $\Gauss$ expansions we are considering.

\section{A Telescopic Nested Summation Algorithm}\label{sec:Algorithm}

We begin this section by outlining the main idea behind our nested summation algorithm. The reader interested in its statement and the outline of its proof may jump directly to subsections \ref{subsec:AlgoStatement} and \ref{subsec:proof}, respectively.

\subsection{Strategy of the Algorithm}
\label{sec:algorithm_strategy}

Let $\Phi(\a| x)$ be a sum of depth $d$, which depends on a continuous variable $x \in \mathbb{C}$ and a symbolic, non-negative integer $\a$. As is typical in telescoping algorithms, our objective is to find a systematic cancellation between the terms in this sum that either decreases the summation range or the sum depth.  

Inspired by the telescoping identity \eqref{SimpleTele}, we begin by writing the following ansatz for this sum:
\begin{align}
	\label{AnsatzTele}
	\Phi(\a| x) = \sum_{\mu=1}^{\a-1} \big[ \Phi(\mu{+}1 | x) - x^P \Phi(\mu| x) \big] x^{Q(\mu)} + \Phi(1| x) x^{R},
\end{align}
where $P$ is known while $Q(\mu)$ and $R$ are to be determined.
Expanding this ansatz and inspecting the coefficients of $\Phi(\mu|x)$ for different values of $\mu$, we see that we must have 
\begin{align} \label{Req1}
Q(1) &= R - P \,, \\
Q(\mu) &= P+Q(\mu-1) \qquad \text{for} \qquad 1 < \mu < \alpha - 1\, , \\
Q(\a-1) &= 0 \, ,
\end{align}
for~\eqref{AnsatzTele} to be consistent. Solving these constraints and plugging the solution back into~\eqref{AnsatzTele}, we arrive at
\begin{align}
	\label{GeneralTele}
	\Phi(\a|x) = \sum_{\mu=1}^{\a-1} \Delta \Phi(\mu|x) x^{(\a-\mu-1)P} + \Phi(1|x) x^{(\a-1)P}
\end{align}
where we have defined
\begin{align} \label{eq:delta_S}
	\Delta \Phi(\mu|x) = \Phi(\mu{+}1|x) - x^P \Phi(\mu|x) \, .
\end{align}
The form of~\eqref{GeneralTele} is analogous to equation~(17) in \cite{Anzai:2012xw}, albeit much less general. However, what we lose in generality we gain in simplicity and computational efficiency.

The strategy is then to determine whether $\Delta \Phi(\mu|x)$ can be reduced to simpler sums than the original $\Phi(\a|x)$. This requires analyzing the specific form of the sums under consideration; in many cases, no reduction will occur (although in some of these cases, a different form of the generalized telescoping identity may work). In the class of sums we focus on in this paper, we will see that the sum $\Delta \Phi(\mu|x)$ has lower depth than $\Phi(\mu|x)$, allowing us to telescopically recurse. We now turn to that analysis.

\subsection{Statement of the Recursion}\label{subsec:AlgoStatement}

Having motivated sums of the form~\eqref{eq:SpecialGaussSum} by considering the expansion of Gauss hypergeometric functions, we now broaden the scope of our analysis to the more general class of sums\footnote{We recall that, in accordance with the conventions established in subsection~\ref{subsec:Zsums}, boldface indices are multi-indices, and the first index has been dropped from primed multi-indices; see in particular the discussion under equation~\eqref{eq:ZsumAbbrev}.}
\begin{equation} \label{eq:general_sums_class}
	{\cal S}^{p,q}_{\mathbf{m};\mathbf{r}}(\a,N|x;\mathbf{y};\mathbf{z}) \equiv \sum_{n=1}^{N} \frac{x^n}{n^p(n+\a)^q} Z_{\mathbf{m}}(n{-}1|\mathbf{y}) Z_{\mathbf{r}}(n{+}\a{-}1|\mathbf{z}) \, ,
\end{equation}
where $\{x, x_1, \dots, x_d, y_1, \dots, y_h\} \in \mathbb{C}$ and $\{N,\a\} \in \mathbb{Z}_{\geq 0}$ are all allowed to be symbolic. We here present an algorithm for converting any sum ${\cal S}^{p,q}_{\mathbf{m};\mathbf{r}}(\a,N|x;\mathbf{y};\mathbf{z})$ with fixed values for $\{p, q\} \in \mathbb{Z}_{\geq 0}$ and $\{m_1,\dots,m_d, r_1,\dots, r_h\} \in \mathbb{Z}_+$ into nested sums that depend on the remaining symbolic parameters.

Our algorithm takes the form of a recursion, in which the application of the relation
\begin{align}
	{\cal S}^{p,q}_{\mathbf{m};\mathbf{r}}(\a,N|x;\mathbf{y};\mathbf{z})  &= x^{-\a}  \sum_{\mu=2}^{\a} \sum_{i=0}^{q-1}  \frac{\binom{p-1+i}{p-1}}{(-1)^{p+1}\a^{p+i}} x^\mu \, {\cal S}^{m_1,q-i}_{\mathbf{m'};\mathbf{r}}(\mu,N|xy_1;\mathbf{y'};\mathbf{z}) \nonumber  \\
	&\qquad \qquad + \sum_{\mu=1}^{\a-1}\sum_{i=0}^{p-1} \frac{\binom{q-1+i}{q-1}}{(-1)^i \a^{q+i}} \, z_1^\mu {\cal S}^{p-i,r_1}_{\mathbf{m};\mathbf{r'}}(\mu,N|x z_1;\mathbf{y};\mathbf{z'}) \label{eq:Spq_recursion} \\
	&\qquad \qquad + {\cal V}^{p,q}_{\mathbf{m};\mathbf{r}}(\a,N|x;\mathbf{y};\mathbf{z}) \, \nonumber
\end{align} 
decreases the overall depth of the $Z$-sums appearing on the right-hand side of equation~\eqref{eq:general_sums_class} in each iteration. The boundary term
\begin{align} 
	{\cal V}^{p,q}_{\mathbf{m};\mathbf{r}}(\a,N|x;\mathbf{y};\mathbf{z}) &\equiv  
	\sum_{i=0}^{p-1} \frac{\binom{q-1+i}{q-1}}{ (-1)^i \a^{q+i}}  {\cal S}^{p-i,0}_{\mathbf{m};\mathbf{r}}(1,N|x;\mathbf{y};\mathbf{z}) \nonumber \\
	& \qquad + x^{-\a} \sum_{i=0}^{q-1} \frac{\binom{p-1+i}{p-1}}{(-1)^p\a^{p+i}} 
	\Bigg[ x \, {\cal S}^{0,q-i}_{\mathbf{m};\mathbf{r}}(1,N|x;\mathbf{y};\mathbf{z}) \label{eq:Spq_recursion_boundary} \\
	& \qquad \qquad \qquad - \delta_{|\mathbf{m}|,0} \Big( Z_{q-i,\mathbf{r}}(\alpha|x,\mathbf{z})  - x \delta_{|\mathbf{r}|,0} \Big) \nonumber \\
	& \qquad \qquad \qquad + Z_{\mathbf{m}}(N|\mathbf{y}) \Big(Z_{q-i,\mathbf{r}}(N{+}\a|x,\mathbf{z}) - Z_{q-i,\mathbf{r}}(N{+}1|x,\mathbf{z}) \Big) \Bigg] \, \nonumber
\end{align}
appears at each step, but can be converted into $Z$-sums using the techniques of~\cite{Moch:2001zr}. Equation~\eqref{eq:Spq_recursion} can be applied until the depth of both $Z$-sums are zero (using the prescription that $Z$-sums of negative depth vanish), whereby we are left with the sum
\begin{align}\label{eq:SpqTermination}
	{\cal S}^{p,q}_{\emptyset;\emptyset}(\a,N|x;\emptyset;\emptyset) &=
	\sum_{i=0}^{p-1} \binom{q{-}1{+}i}{q{-}1} \frac{(-1)^i}{\a^{q+i}} Z_{p-i}(N|x) \\ & 
	\qquad \qquad + x^{-\a} \sum_{i=0}^{q-1} \binom{p{-}1{+}i}{p{-}1} \frac{(-1)^p}{\a^{p+i}}
	\Big[ Z_{q-i}(N{+}\a|x) - Z_{q-i}(\a|x) \Big] \, . \nonumber
\end{align}
We then proceed to convert the sums over $\mu$ that appeared at each step in the recursion into nested sums. When $N$ is taken to be infinite, these sums can be evaluated as $Z$-sums, again using the methods of~\cite{Moch:2001zr}. For generic finite $N$, these sums evaluate to the more general class of cyclotomic harmonic sums~\cite{Ablinger:2011te}. Once these sums have been converted into nested sums, we are left with sums that can be carried out explicitly for fixed values of $p$ and $q$. Altogether, this recursion converts the sum over $n$ in~\eqref{eq:general_sums_class} into a linear combination of $Z$-sums or cyclotomic harmonic sums with coefficients that depend on $\alpha$, $N$, $x$, $\mathbf{y}$, and $\mathbf{z}$.

In the rest of this section, as well as in sections~\ref{sec:double_pentaladder} and~\ref{sec:self_energy}, we will focus on sums for which $N$ is infinite; we provide more details for sums with generic $N$ in the appendix. For the cases we focus on here, we abbreviate
\begin{align}
{\cal S}^{p,q}_{\mathbf{m};\mathbf{r}}(\a|x;\mathbf{y};\mathbf{z}) &\equiv \lim_{N\to \infty} {\cal S}^{p,q}_{\mathbf{m};\mathbf{r}}(\a,N|x;\mathbf{y};\mathbf{z}) \, , \\
{\cal V}^{p,q}_{\mathbf{m};\mathbf{r}}(\a|x;\mathbf{y};\mathbf{z})  &\equiv \lim_{N\to \infty} {\cal V}^{p,q}_{\mathbf{m};\mathbf{r}}(\a,N|x;\mathbf{y};\mathbf{z}) \, .
\end{align}
A number of simplifications occur in this limit. For instance, the last line in~\eqref{eq:Spq_recursion_boundary} vanishes, and many of the $Z$-sums that appear can be converted into generalized polylogarithms using~\eqref{eq:Z_sums_to_polylogs}. However, there always remain $Z$-sums that cannot be expressed as polylogarithms, as their upper summation bound only depends on $\a$. This can be seen, for instance, in the terminal sum~\eqref{eq:SpqTermination}, which becomes 
\begin{align} \label{eq:SpqTermination_infinite}
	{\cal S}^{p,q}_{\emptyset;\emptyset}(\a|x;\emptyset;\emptyset)  &=
	\sum_{i=0}^{p-1} \binom{q{-}1{+}i}{q{-}1} \frac{(-1)^i}{\a^{q+i}} \li_{p-i}(x) \\ & 
	\qquad \qquad + x^{-\a} \sum_{i=0}^{q-1} \binom{p{-}1{+}i}{p{-}1} \frac{(-1)^p}{\a^{p+i}}
	\Big[ \li_{q-i}(x) - Z_{q-i}(\a|x) \Big] \,  \nonumber
\end{align}
in this limit. For any specific choice of $\a$, these remaining $Z$-sums evaluate to rational functions of their arguments.

\paragraph{A Simple Illustration of the Algorithm}~\\[-10pt] 

\noindent Let us work though a example to see how this recursion works in practice. Consider the sum ${\cal S}^{0,1}_{1;1}(\a|x;y_1;z_1)$. Iteratively applying~\eqref{eq:Spq_recursion}, we get
\begin{align}
	{\cal S}^{0,1}_{1;1}(\a|x;y_1;z_1)  &= - x^{-\a}  \sum_{\mu=2}^{\a} x^\mu \, {\cal S}^{1,1}_{\emptyset;1}(\mu|x y_1;\emptyset;z_1) + {\cal V}^{0,1}_{1;1}(\a|x;y_1;z_1) \, . \label{eq:example_eq_1} 
\end{align}
and then
\begin{align}
{\cal S}^{1,1}_{\emptyset;1}(\mu|x y_1;\emptyset;z_1) &= \frac{1}{\mu}  \sum_{\nu=1}^{\mu-1}  \, z_1^\nu {\cal S}^{1,1}_{\emptyset;\emptyset}(\nu|x y_1 z_1;\emptyset;\emptyset) + {\cal V}^{1,1}_{\emptyset;1}(\mu|xy_1;\emptyset;z_1) \, .   \label{eq:example_eq_2}  
\end{align}
At this point, the recursion terminates in an expression of the form~\eqref{eq:SpqTermination_infinite}, and we need to convert the sums over $\nu$ and $\mu$ into $Z$-sums. In general, this involves some combination of partial fractioning, reindexing sums, and applying~\eqref{ZIdentity} to shift the upper bound of existing $Z$-sums. Applying these techniques, it is not hard to show that
\begin{align}
{\cal S}^{1,1}_{\emptyset;1}(\mu|x y_1;\emptyset;z_1) &= \frac{1}{\mu} \Big(Z_1(\mu{-}1|z_1) - Z_1\big(\mu{-}1\big|{\textstyle \frac{1}{x y_1}}\big) \Big) \li_1(x y_1 z_1) \nonumber \\
&\qquad \qquad + \frac{1}{\mu} Z_{1,1}\big(\mu{-}1 \big| {\textstyle \frac{1}{x y_1}}, x y_1 z_1\big) + \frac{1}{\mu} Z_{2}(\mu{-}1| z_1) \label{eq:example_eq_3}  \\
&\qquad \qquad +  {\cal V}^{1,1}_{\emptyset;1}(\mu|xy_1;\emptyset;z_1) \, , \nonumber
\end{align}
and that the boundary contribution is given by
\begin{align}
{\cal V}^{1,1}_{\emptyset;1}(\mu|xy_1;\emptyset;z_1) &= \frac{1}{\mu} \big( \li_{1,1}(z_1,x y_1) + \li_2(x y_1 z_1) \big)  \\
&\qquad \qquad - \frac{(x y_1)^{-\mu}}{\mu} \big( \li_{1,1}(z_1, x y_1) - Z_{1,1}( \mu | x y_1, z_1)  \big) \, . \nonumber
\end{align}
Using these results, the sum over $\mu$ in~\eqref{eq:example_eq_1} can be carried out using the same techniques, whereby we find 
\begin{align}
{\cal S}^{0,1}_{1;1}(\a|x;y_1;z_1)  &= - x^{-\a} \bigg( \left(Z_{1,1}(\a|x,z_1) - Z_{1,1}\big(\a \big|x,{\textstyle \frac{1}{x y_1}}\big)\right) \li_1(x y_1 z_1) \nonumber\\
&\qquad \qquad +  Z_{1,1,1}\big(\a| x, {\textstyle \frac{1}{x y_1}}, x y_1 z_1\big) + Z_{1,2}(\a | x, z_1)   \nonumber\\
&\qquad \qquad + \big( Z_1(\a|x) - x \big) \big( \li_{1,1}(z_1,x y_1) + \li_2(x y_1 z_1) \big) \label{eq:example_eq_4} \\
&\qquad \qquad - \left ( Z_1\big(\a\big|{\textstyle \frac{1}{y_1}}\big) - \frac{1}{y_1} \right) \li_{1,1}(z_1, x y_1) \nonumber \\
&\qquad \qquad +  Z_{1,1,1}\big( \a \big | {\textstyle \frac{1}{y_1}}, x y_1, z_1 \big) + Z_{2,1} (\a | x, z_1 ) \bigg)  \nonumber \\
&\qquad \qquad + {\cal V}^{0,1}_{1;1}(\a|x;y_1;z_1) \, . \nonumber
\end{align}
To evaluate the boundary contribution ${\cal V}^{0,1}_{1;1}(\a|x;y_1;z_1)$ we need to be slightly more careful. Evaluating the sums over $i$ in~\eqref{eq:Spq_recursion_boundary} for these indices and arguments, we find
\begin{align}
{\cal V}^{0,1}_{1;1}(\a|x;y_1;z_1) = x^{-\a} \sum_{n=2}^\infty \frac{x^{n}}{n} Z_1(n-2|y_1) Z_1(n-1| z_1) \, , \label{eq:example_eq_5}
\end{align}
where we have additionally shifted the summation index $n \to n{-}1$ to put the denominator in a form that fits the definition of $Z$-sums. Since the summand on the right hand side is zero when $n=1$, we can change the lower summation bound back to $1$. To increase the upper summation bound of $Z_{\mathbf{m}}(n{-}2)$, we would like to use equation~\eqref{ZIdentity}; however, we need to emend this relation so that it remains valid when $n=1$. Doing so, we have
\begin{align} \label{eq:Z_arg_shift}
Z_{\mathbf{m}}(n{-}2|\mathbf{y}) = Z_{\mathbf{m}}(n{-}1|\mathbf{y}) - \frac{y_1^{n-1}}{(n-1)^{m_1}}  Z_{\mathbf{m'}}(n{-}2|\mathbf{y'}) \theta(n-2) - \delta_{|m|,0} \delta_{n,1} \, . 
\end{align}
where $\theta(k)$ is the Heaviside function, which is equal to $1$ when $k\ge0$ and 0 otherwise. The Heaviside function makes clear that this term must vanish when $n=1$ (since all other terms in~\eqref{eq:Z_arg_shift} vanish), despite the ambiguity of both its numerator and denominator evaluating to zero. Substituting this relation into~\eqref{eq:example_eq_5} and converting the expression into $Z$-sums like above, one finds
\begin{align}
{\cal V}^{0,1}_{1;1}(\a|x;y_1;z_1) &= x^{-\a} \bigg( \li_{2,1}(y_1 z_1, x) + \li_{1,1,1} (y_1,z_1,x) + \li_{1,1,1}(z_1,y_1,x) \\
&\qquad \qquad - x \li_2(x y_1 z_1) - \left( x - \frac{1}{y_1} \right) \li_{1,1}(z_1, x y_1) \bigg) \nonumber \, .
\end{align} 
Putting this all together, we find
\begin{align}
{\cal S}^{0,1}_{1;1}(\a|x;y_1;z_1)  &= x^{-\a} \bigg( \li_{1,1,1} (y_1,z_1,x) + \li_{1,1,1}(z_1,y_1,x) + \li_{2,1}(y_1 z_1, x) \nonumber\\
&\qquad \qquad + \li_{1,1}(z_1, x y_1)\left( Z_1\big(\a\big|{\textstyle \frac{1}{y_1}}\big) - Z_1(\a|x) \right) - \li_2(x y_1 z_1) Z_1(\a|x)  \nonumber  \\
&\qquad \qquad + \li_1(x y_1 z_1) \left(Z_{1,1}\big(\a \big|x,{\textstyle \frac{1}{x y_1}}\big) - Z_{1,1}(\a|x,z_1)\right) \label{eq:example_eq_6}\\
&\qquad \qquad -  Z_{1,1,1}\big(\a| x, {\textstyle \frac{1}{x y_1}}, x y_1 z_1\big) - Z_{1,2}(\a | x, z_1)   \nonumber\\
&\qquad \qquad -  Z_{1,1,1}\big( \a \big | {\textstyle \frac{1}{y_1}}, x y_1, z_1 \big) - Z_{2,1} (\a | x, z_1 )  \bigg) \nonumber \, .
\end{align}
As emphasized above, this expression depends on not only polylogarithms but also $Z$-sums with upper summation bound $\a$. However, for any specific value of $\a$, this expression reduces to a linear combination of generalized polylogarithms with rational coefficients that depend on $x$, $y_1$, and $z_1$. 

More complicated sums of this class can be evaluated following the same strategy, but require an increasing amount of algebra and quickly become tedious. In the ancillary files included with this paper, we provide a {\sc Mathematica} package that can be used to apply this algorithm to any sum in the class~\eqref{eq:general_sums_class}, up to computer memory and time limitations.

As we also mentioned in the introduction, let us remind the reader that for specific values of $\mathbf{m}, \mathbf{r}, p$ and $q$ in equation~\eqref{eq:general_sums_class}, a recurrence in $N$ may also be found and solved with the help of the symbolic summation algorithms implemented in the package {\sc {Sigma}} \cite{article}. Nevertheless, we notice that even for ${\cal S}^{1,1}_{1;1}(\a|x;y_1;z_1)$, which is only slightly more complicated than the example of eq.~\eqref{eq:example_eq_6}, the same package produces a recurrence in $\alpha$ that contains tens of thousands of terms, which cannot be solved in a reasonable amount of time. It is in this sense, that our algorithm provides a novel, alternative route to the computations of this class of sums.

\subsection{Proof for Euler-Zagier Sums}
\label{subsec:proof}

In this section we illustrate the derivation of~\eqref{eq:Spq_recursion} by proving it in the $N \to \infty$ limit for the case of Euler-Zagier sums, which corresponds to the restriction $\mathbf{y} = 1,\dots,1$ and $\mathbf{z} = 1,\dots,1$ in~\eqref{eq:general_sums_class}. This simplifies the notational clutter without altering the telescopic strategy, which can also be applied in the general case. We prove the more general result in appendix~\ref{appendix:general_telescoping_algorithm}. 

We abbreviate the sums we focus on in this section by 
\begin{align} 
{\cal S}^{p,q}_{\mathbf{m};\mathbf{r}}(\a|x) &\equiv {\cal S}^{p,q}_{\mathbf{m};\mathbf{r}}(\a|x;1,\dots,1;1,\dots,1) \\
&= \sum_{n=1}^{\infty} \frac{x^n}{n^p(n+\a)^q} Z_{\mathbf{m}}(n{-}1) Z_{\mathbf{r}}(n{+}\a{-}1) \, . \label{eq:Spq_EZ_def}
\end{align}
A generic sum of this form can be split into a linear combination of cases in which either $p$ or $q$ is zero by partial fractioning: 
\begin{equation} \label{eq:partial_fractioned_Spq}
	{\cal S}^{p,q}_{\mathbf{m};\mathbf{r}}(\a|x) = 
	\sum_{i=0}^{p-1}  \frac{\binom{q-1+i}{q-1}}{(-1)^i \a^{q+i}} {\cal S}^{p-i,0}_{\mathbf{m};\mathbf{r}}(\a|x) +  
	\sum_{i=0}^{q-1}  \frac{\binom{p-1+i}{p-1}}{(-1)^p \a^{p+i}} {\cal S}^{0,q-i}_{\mathbf{m};\mathbf{r}}(\a|x) \, .
\end{equation}
Our strategy will be to find telescopic recursions for ${\cal S}^{p,0}_{\mathbf{m};\mathbf{r}}(\a|x)$ and ${\cal S}^{0,q}_{\mathbf{m};\mathbf{r}}(\a|x)$ separately, after which the case of general $p$ and $q$ can be treated by plugging these recursive formulas into~\eqref{eq:partial_fractioned_Spq}.

\paragraph{Telescoping \texorpdfstring{${\cal S}^{p,0}_{\mathbf{m};\mathbf{r}}(\a|x)$}{Sp0}}~\\[-10pt] 

\noindent To find a telescopic recursion on ${\cal S}^{p,0}_{\mathbf{m};\mathbf{r}}(\a|x)$, we consider the shifted sum
\begin{align}
	{\cal S}^{p,0}_{\mathbf{m};\mathbf{r}}(\a+1|x) &=  
	\sum_{n=1}^{\infty} \frac{x^n}{n^p} Z_{\mathbf{m}}(n{-}1) Z_{\mathbf{r}}(n{+}\a) \\
	&= 	\sum_{n=1}^{\infty} \frac{x^n}{n^p} Z_{\mathbf{m}}(n{-}1) Z_{\mathbf{r}}(n{+}\a {-} 1)  \\
	&\qquad \qquad + \sum_{n=1}^{\infty} \frac{x^n}{n^p (n+\a)^{r_1}} Z_{\mathbf{m}}(n{-}1) Z_{\mathbf{r'}}(n{+}\a {-} 1) \nonumber \\
	&= 	{\cal S}^{p,0}_{\mathbf{m};\mathbf{r}}(\a|x) + {\cal S}^{p,r_1}_{\mathbf{m};\mathbf{r'}}(\a|x) \, , \label{eq:Sp0_alphaPlusOne}
\end{align}
where in the second line we have used identity~\eqref{ZIdentity} to shift the upper summation index of $Z_{\mathbf{r}}(n+\a)$. In principle, we should treat the $|\mathbf{r}| = 0$ case separately, since~\eqref{ZIdentity} can't be applied to depth-zero sums; however, it is easy to see that ${\cal S}^{p,0}_{\mathbf{m};\emptyset}(\a|x)$ is independent of $\alpha$, and that~\eqref{eq:Sp0_alphaPlusOne} correspondingly gives the correct answer as long as we adopt the prescription that $Z$-sums with negative depth evaluate to zero.

Comparing to~\eqref{eq:delta_S}, we see that the parameter $P$ in our ansatz is in this case zero, and we have
\begin{align}
\Delta {\cal S}^{p,0}_{\mathbf{m};\mathbf{r}}(\a|x) &\equiv {\cal S}^{p,0}_{\mathbf{m};\mathbf{r}}(\a+1|x) - {\cal S}^{p,0}_{\mathbf{m};\mathbf{r}}(\a|x) \\
&=  {\cal S}^{p,r_1}_{\mathbf{m};\mathbf{r'}}(\a|x) \, .
\end{align}
Thus, plugging this difference into~\eqref{GeneralTele}, we obtain
\begin{align}	
	\label{eq:Sp0_recursion}
	 {\cal S}^{p,0}_{\mathbf{m};\mathbf{r}}(\a|x) =
	\sum_{\mu=1}^{\a-1} {\cal S}^{p,r_1}_{\mathbf{m};\mathbf{r'}}(\mu|x) + {\cal S}^{p,0}_{\mathbf{m};\mathbf{r}}(1|x) \, .
\end{align}
Importantly, the sums on the right hand side of this equation are all strictly simpler than the original; the terms in the sum over $\mu$ all involve a $Z$-sum of one lower depth, and the last term does not depend on $\alpha$. Note, however, that the terms in the sum over $\mu$ no longer satisfy $q=0$; thus, while we can again split these sums up into cases in which either $p$ or $q$ is zero as in~\eqref{eq:partial_fractioned_Spq}, we will need to be able to reduce the depth of sums with nonzero $q$ in order to achieve a genuine recursion in depth.  

\paragraph{Telescoping \texorpdfstring{${\cal S}^{0,q}_{\mathbf{m};\mathbf{r}}(\a|x)$}{S0q}}~\\[-10pt] 

\noindent Following the same strategy as above, we consider
\begin{align} 
	{\cal S}^{0,q}_{\mathbf{m};\mathbf{r}}(\a+1|x) &= \sum_{n=2}^{\infty} \frac{x^{n-1}}{(n+\a)^q} Z_{\mathbf{m}}(n{-}2) Z_{\mathbf{r}}(n{+}\a{-}1)\, , \label{eq:S0q_shifted}
\end{align}
where we have shifted the summation index by $n \to n {-}1$ relative to the definition~\eqref{eq:Spq_EZ_def}. Since the summand on the right hand side of~\eqref{eq:S0q_shifted} is zero when $n=1$, we can change the lower summation bound back to $1$. To increase the upper summation bound of $Z_{\mathbf{m}}(n{-}2)$, we again use equation~\eqref{eq:Z_arg_shift}. This gives us
\begin{align}
{\cal S}^{0,q}_{\mathbf{m};\mathbf{r}}(\a+1|x) &= \sum_{n=1}^{\infty} \frac{x^{n-1}}{(n+\a)^q} Z_{\mathbf{m}}(n{-}1) Z_{\mathbf{r}}(n{+}\a{-}1) - \frac{\delta_{|\mathbf{m}|,0} }{(\a+1)^q} Z_{\mathbf{r}}(\a)  \\
	&\qquad \qquad - \sum_{n=1}^{\infty} \frac{x^{n}}{ n^{m_1}(n+\a + 1)^q} Z_{\mathbf{m'}}(n{-}1) Z_{\mathbf{r}}(n{+}\a) \nonumber \\ 
	&= \frac{1}{x} {\cal S}^{0,q}_{\mathbf{m};\mathbf{r}}(\a|x) - {\cal S}^{m_1,q}_{\mathbf{m'};\mathbf{r}}(\a+1|x)- \frac{\delta_{|\mathbf{m}|,0} }{(\a+1)^q} Z_{\mathbf{r}}(\a)  \, .
\end{align}
where we have shifted $n \to n{+}1$ in the sum involving the Heaviside function to get the expression into this form. 

Comparing to~\eqref{eq:delta_S}, we see that $P={-}1$ and that
\begin{align}
	\Delta {\cal S}^{0,q}_{\mathbf{m};\mathbf{r}}(\a|x) &\equiv {\cal S}^{0,q}_{\mathbf{m};\mathbf{r}}(\a+1|x) - \frac{1}{x} {\cal S}^{0,q}_{\mathbf{m};\mathbf{r}}(\a|x) \\
	&= - {\cal S}^{m_1,q}_{\mathbf{m'};\mathbf{r}}(\a+1|x)  - \frac{\delta_{|\mathbf{m}|,0}}{(\a+1)^q} Z_{\mathbf{r}}(\a)  \, .
	\end{align}
Therefore, our ansatz~\eqref{GeneralTele} gives the relation
\begin{align}
	\label{eq:S0q_recursion}
	{\cal S}^{0,q}_{\mathbf{m};\mathbf{r}}(\a|x) &= 
	- \sum_{\mu=1}^{\a-1} x^{\mu-\a+1} {\cal S}^{m_1,q}_{\mathbf{m'};\mathbf{r}}(\mu+1|x)+ x^{1-\a}  {\cal S}^{0,q}_{\mathbf{m};\mathbf{r}}(1|x)    \\	
	&\qquad \qquad - x^{-\a} \delta_{|\mathbf{m}|,0} \Big( Z_{q,\mathbf{r}}(\alpha|x, 1,\dots,1)  - x \delta_{|\mathbf{r}|,0} \Big)  \, , \nonumber
	\end{align}
where we have already converted one of the sums over $\mu$ into a $Z$-sum. As in~\eqref{eq:Sp0_recursion}, all the terms on the right hand side are simpler than the original sum; the sums in the first line either have one lower depth or don't depend on $\a$, and the terms in the second line no longer involve a sum over $n$.

\newpage
\paragraph{The Closed Recursion for \texorpdfstring{${\cal S}^{p,q}_{\mathbf{m};\mathbf{r}}(\a|x) $}{Spq}}~\\[-10pt] 

\noindent Substituting equations \eqref{eq:Sp0_recursion} and \eqref{eq:S0q_recursion} into \eqref{eq:partial_fractioned_Spq}, we obtain the recursion relation
\begin{align}
	\label{eq:EZ_Spq_recursion}
	{\cal S}^{p,q}_{\mathbf{m};\mathbf{r}}(\a|x)  &= x^{-\a}  \sum_{\mu=2}^{\a} \sum_{i=0}^{q-1}  \frac{\binom{p-1+i}{p-1}}{(-1)^{p+1} \a^{p+i}} x^\mu \, {\cal S}^{m_1,q-i}_{\mathbf{m'};\mathbf{r}}(\mu|x)  \nonumber \\
		&\qquad \qquad + \sum_{\mu=1}^{\a-1}\sum_{i=0}^{p-1} \frac{\binom{q-1+i}{q-1}}{(-1)^i \a^{q+i}} \, {\cal S}^{p-i,r_1}_{\mathbf{m};\mathbf{r'}}(\mu|x)  +{\cal V}^{p,q}_{\mathbf{m};\mathbf{r}}(\a|x) 
\end{align}
which is the simplified version of~\eqref{eq:Spq_recursion} that applies to Euler-Zagier sums in cases where $N \to \infty$. We have collected all the boundary terms into
\begin{align}  \label{eq:EZ_Spq_recursion_boundary} 
{\cal V}^{p,q}_{\mathbf{m};\mathbf{r}}(\a|x) &\equiv  \sum_{i=0}^{p-1} \frac{\binom{q-1+i}{q-1}}{ (-1)^i \a^{q+i}}  {\cal S}^{p-i,0}_{\mathbf{m};\mathbf{r}}(1|x) + x^{-\a} \sum_{i=0}^{q-1}  \frac{\binom{p-1+i}{p-1}}{(-1)^p \a^{p+i}} \Bigg[ x \,  {\cal S}^{0,q-i}_{\mathbf{m};\mathbf{r}}(1|x) \\
&\qquad \qquad  - \delta_{|\mathbf{m}|,0} \left( Z_{q-i,\mathbf{r}}(\alpha|x, 1,\dots,1)  -x \delta_{|\mathbf{r}|,0}\right) \Bigg] \, . \nonumber 
\end{align}
Like the more general recursion~\eqref{eq:Spq_recursion}, the relation~\eqref{eq:EZ_Spq_recursion} can be applied iteratively until the depth of both $Z$-sums are zero, terminating in the sum~\eqref{eq:SpqTermination}. 

\section{Application I: The Double Pentaladder Integrals \texorpdfstring{$\Omega^{(L)}$}{Omega}} \label{sec:double_pentaladder}

\begin{figure}[t]
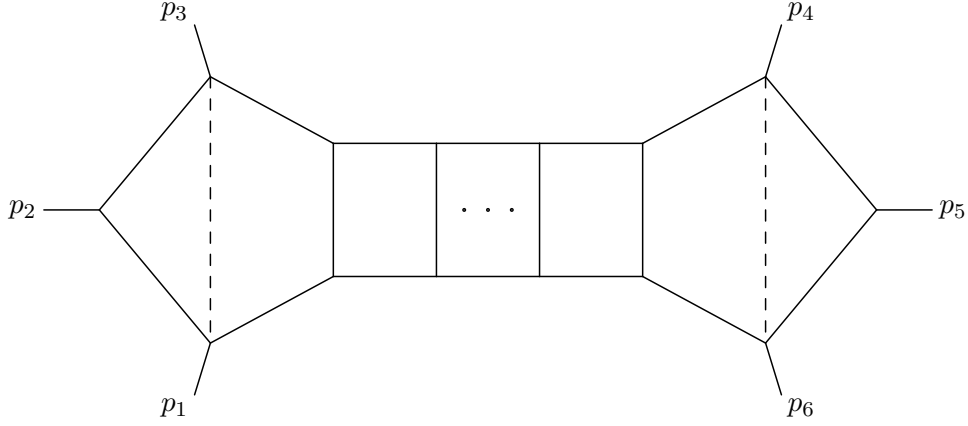

\begin{center}
\begin{fmfchar*}(300,140)
	\fmfset{dot_size}{0.18mm}
	 %
	\fmfforce{(.01w,.5h)}{p1}
	\fmfforce{(.15w,.86h)}{p2}
	\fmfforce{(.305w,.68h)}{p3}
	\fmfforce{(.435w,.68h)}{p4}
	\fmfforce{(.565w,.68h)}{p5}
	\fmfforce{(.695w,.68h)}{p6}
	\fmfforce{(.85w,.86h)}{p7}
	\fmfforce{(.99w,.5h)}{p8}
	\fmfforce{(.85w,.14h)}{p9}
	\fmfforce{(.695w,.32h)}{p10}
	\fmfforce{(.565w,.32h)}{p11}
	\fmfforce{(.435w,.32h)}{p12}
	\fmfforce{(.305w,.32h)}{p13}
	\fmfforce{(.15w,.14h)}{p14}
	%
	\fmfforce{(.47w,.5h)}{d1}
	\fmfforce{(.5w,.5h)}{d2}
	\fmfforce{(.53w,.5h)}{d3}
	%
	\fmfforce{(-.06w,.5h)}{e1}
	\fmfforce{(.13w,1.0h)}{e2}
	\fmfforce{(.87w,1.0h)}{e3}
	\fmfforce{(1.06w,.5h)}{e4}
	\fmfforce{(.87w,0.0h)}{e5}
	\fmfforce{(.13w,0.0h)}{e6}
	%
	\fmfforce{(.02w,.84h)}{l1}
	\fmfforce{(.5w,.95h)}{l2}
	\fmfforce{(.98w,.84h)}{l3}
	\fmfforce{(.98w,.16h)}{l4}
	\fmfforce{(.5w,.05h)}{l5}
	\fmfforce{(.02w,.16h)}{l6}
	%
	\fmf{plain, width=.2mm}{p1,p2}
	\fmf{plain, width=.2mm}{p2,p3}
	\fmf{plain, width=.2mm}{p3,p4}
	\fmf{plain, width=.2mm}{p4,p5}
	\fmf{plain, width=.2mm}{p5,p6}
	\fmf{plain, width=.2mm}{p6,p7}
	\fmf{plain, width=.2mm}{p7,p8}
	\fmf{plain, width=.2mm}{p8,p9}
	\fmf{plain, width=.2mm}{p9,p10}
	\fmf{plain, width=.2mm}{p10,p11}
	\fmf{plain, width=.2mm}{p11,p12}
	\fmf{plain, width=.2mm}{p12,p13}
	\fmf{plain, width=.2mm}{p13,p14}
	\fmf{plain, width=.2mm}{p14,p1}
	%
	\fmf{dashes, width=.2mm}{p2,p14}
	\fmf{dashes, width=.2mm}{p7,p9}
	%
	\fmf{plain, width=.2mm}{p3,p13}
	\fmf{plain, width=.2mm}{p4,p12}
	\fmf{plain, width=.2mm}{p5,p11}
	\fmf{plain, width=.2mm}{p6,p10}
	%
	\fmf{plain, width=.2mm}{p1,e1}
	\fmf{plain, width=.2mm}{p2,e2}
	\fmf{plain, width=.2mm}{p7,e3}
	\fmf{plain, width=.2mm}{p8,e4}
	\fmf{plain, width=.2mm}{p9,e5}
	\fmf{plain, width=.2mm}{p14,e6}
	%
	\fmfdot{d1}
	\fmfdot{d2}
	\fmfdot{d3}
	%
	\fmfv{label={$p_2$}, label.dist=.1cm}{e1}
	\fmfv{label={$p_3$}, label.dist=.1cm}{e2}
	\fmfv{label={$p_4$}, label.dist=.1cm}{e3}
	\fmfv{label={$p_5$}, label.dist=.1cm}{e4}
	\fmfv{label={$p_6$}, label.dist=.1cm}{e5}
	\fmfv{label={$p_1$}, label.dist=.1cm}{e6}
	\end{fmfchar*}
\end{center}
\caption{The double pentaladder integral $\Omega^{(L)}$, built out of an $(L{-}2)$-loop box ladder capped on each end by a pentagon loop. The dashed lines represent specific numerator factors.}
\label{fig:omega_L}
\end{figure}

Let us now explore the power of our algorithm by considering some applications. Our first example is a family of integrals first introduced in \cite{ArkaniHamed:2010kv} in the context of $\mathcal{N}=4$ supersymmetric Yang-Mills theory---the double pentaladder, or $\Omega^{(L)}$,  integrals. These integrals consist of $(L{-}2)$-loop box ladders capped on either end by a pentagon loop that comes with a numerator factor. These diagrams are depicted in figure~\ref{fig:omega_L}. Their numerator renders them infrared finite and parity even, and they depend on the kinematic variables
\begin{align}
x &= 1+\frac{1-u-v-w+\sqrt{\Delta}}{2uv}
    \,, \label{defx}\\
y &= 1+\frac{1-u-v-w-\sqrt{\Delta}}{2uv}
    \,, \label{defy}\\
z &= \frac{u(1-v)}{v(1-u)}
   \,, \label{defz}
\end{align}
where
\be
\Delta=(1-u-v-w)^2-4 u v w
\ee
and
\be
 u=\frac{s_{12}s_{45}}{s_{123}s_{345}},\quad
 v=\frac{s_{23}s_{56}}{s_{234}s_{123}},\quad
 w=\frac{s_{34}s_{61}}{s_{345}s_{234}} \label{uvw_def_app}
\ee
\begin{sloppypar}
\noindent denote the more widely used dual-conformal cross-ratios of the Mandelstam invariants 
$\smash{s_{i\ldots,j}=(p_i+\ldots + p_j)^2}$. 
\end{sloppypar}

In~\cite{Drummond:2010cz, Dixon:2011nj}, it was shown that $\Omega^{(L)}$ is related to $\Omega^{(L-1)}$ by a second-order differential equation. This differential equation was solved for generic values of the coupling $g^2$ in~\cite{Caron-Huot:2018dsv} by
\begin{align} \label{eq:finite_coupling_omega}
 \Omega(x,y,z,g^2)\ &\equiv\ \sum_{L=0}^\infty \,(-g^{2})^L\,\Omega^{(L)} (x,y,z) \\
 &= \int_{-\infty}^\infty \frac{d\nu}{2i} \, z^{i\nu/2} \, 
 \frac{(xy)^{i\nu/2} \mathcal{F}_{\nu}(x)\mathcal{F}_{\nu}(y)-(xy)^{-i\nu/2} \mathcal{F}_{-\nu}(x)\mathcal{F}_{-\nu}(y)}{\sinh(\pi \nu)\
      }\,,
\label{result double penta} 
\end{align}
where
\begin{align} \label{eq:defF} 
 \mathcal{F}_{\nu}(x) &\equiv C(\nu,g) \,{}_2F_1 \left(\frac{i\nu+i\sqrt{\nu^2+4g^2}}{2},\frac{i\nu-i\sqrt{\nu^2+4g^2}}{2},1+i\nu,x \right)
\end{align}
is a hypergeometric function that has been normalized by the factor
\begin{align}
 C(\nu, g) &\equiv \frac{\Gamma\left(1+\frac{i\nu+i\sqrt{\nu^2+4g^2}}{2}\right)\Gamma \left(1+\frac{i\nu-i\sqrt{\nu^2+4g^2}}{2}\right)}{\Gamma(1+i\nu)}\, ,
\end{align}
such that $\mathcal{F}_{\nu}(1) = 1$. Note that $\mathcal{F}_{\nu}(x)$ depends on the coupling $g^2$, although we have left this dependence implicit.

By virtue of Cauchy's residue theorem, the generating function \eqref{result double penta} can put in an equivalent sum representation
\begin{equation}\label{eq:OmegaSum}
	\Omega(x,y,z,g^2) = - \sum_{\a=1}^{\infty} \big[ (-\sqrt{xyz})^\a + (-\sqrt{xy/z})^\a \big] \mathcal{F}_{-i\a}(x)\mathcal{F}_{-i\a}(y) - \mathcal{F}_{0}(x)\mathcal{F}_{0}(y)\, .
\end{equation}
Expanding this expression around small coupling with the use of~\eqref{eq:2F1Expansion}, one can derive a sum representation for each of the perturbative double pentaladder integrals $\Omega^{(L)}$. As noted in~\cite{Caron-Huot:2018dsv}, it is advantageous to carry out this expansion in two steps: first with respect to the small parameter 
\be\label{eq:eps_to_g}
\eps\equiv\frac{\a}{2}\left(\sqrt{1-\frac{4g^2}{\a^2}}-1\right) \, ,
\ee
and then by expanding $\eps$ with respect to $g$. In~\cite{Caron-Huot:2018dsv}, the resulting sum representation for $\Omega^{(L)}$ was evaluated in terms of generalized polylogarithms in the $u \to 1$ and $w \to 0$ limits through eight loops. However, the techniques used there were not sufficiently powerful to evaluate the sum in general kinematics.

Using~\eqref{eq:eps_to_g} to replace $g$ with $\epsilon$, we see that the hypergeometric function in
\be
\mathcal{F}_{-i\alpha}(x)=\frac{\Gamma (1-\epsilon ) \Gamma (1+\alpha +\epsilon)}{\Gamma (1+\alpha )} \Gauss(\alpha+\epsilon,-\epsilon,1+\alpha,x)\,,\label{eq:F_epsilon}
\ee
takes the form~\eqref{eq:2F1_constraint}, and thus falls into the class of hypergeometric expansions our algorithm can handle. Leveraging this fact, we now proceed to evaluate the sum representation of $\Omega^{(L)}$ in general kinematics. 

\subsection{Evaluating the hypergeometric function}\label{subsec:calFExpansion}

We begin by converting the building blocks $\mathcal{F}_{-i\alpha}(x)$ into $Z$-sums. Using the techniques discussed in subsection~\ref{subsec:GaussExp}, it is possible to express these functions as
\be
\mathcal{F}_{-i\alpha}(x) = \frac{\pi \epsilon}{\sin \pi \epsilon}\Biggl[
\sum_{i=0}^\infty \epsilon^i Z_{\fwboxL{27pt}{{\underbrace{1,\ldots,1}_i}}}(\alpha)+ g^2\sum_{i,j=0}^\infty
(-1)^i \epsilon^{i+j} S_{i,j}(\a|x) \Biggr] \,,
\label{eq:Fjk_Expansion}
\ee
where we have denoted the relevant specialization of the sums \eqref{eq:Spq_EZ_def} by
\begin{align}
	\label{Sk}
	S_{i,j}(\a|x) &\equiv {\cal S}^{1,1}_{\ub{1}{i};\ub{1}{j}}(\a|x) \\
	&=  \sum_{n=1}^{\infty} \frac{x^n}{n(n+\a)} Z_{\ub{1}{i}}(n-1) Z_{\ub{1}{j}}(n+\a-1) \label{Sk_2}
\end{align}
in  order to avoid notational clutter. We highlight that $\eps$ in~\eqref{eq:Fjk_Expansion} implicitly depends on $\alpha$ and $g^2$, as per the definition~\eqref{eq:eps_to_g}.

The $\alpha\to 0$ case of~\eqref{eq:Fjk_Expansion} may be obtained by taking the smooth limit $\epsilon\to i g$, as can be seen from~\eqref{eq:eps_to_g}. In this case, the sum over $n$ may be readily evaluated with the techniques of \cite{Moch:2001zr}, where after using stuffle relations to combine the product of $Z$-sums it can be evaluated in terms of HPLs,
\begin{equation}
	\sum_{n=1}^{\infty} \frac{x^n}{n^{m_1}} Z_{m_2,\ldots,m_d}(n-1) = 
	H_{m_d,\ldots,m_1}(x)\, ,
\end{equation}
consistent with equation~\eqref{eq:HPL}.\footnote{
	Incidentally, one can construct a non-recursive version of the relevant stuffle relation as follows. First define the sum over all permuations of weight indices 
	\begin{equation*}
		\sigma(a,b) \equiv \sum_{\sigma \in S_{a+b}}
		Z_{ \sigma [ \ub{1}{a} \ub{2}{b} ] }(n-1).
	\end{equation*}
	The sum over the elements of permuation group $S_{a+b}$ contains $\frac{(a+b)!}{a!b!} \equiv \rho(a,b)$ elements, with each $Z$-sum having weight $a+2b$. Then one has
	\begin{equation*}
		Z_{\ub{1}{i}}(n-1) Z_{\ub{1}{j}}(n-1) = \sum_{k=0}^{\min(i,j)} \rho(i-k,j-k) \sigma(i+j-2k,k) \, .
	\end{equation*}}
In this manner we observe that
\begin{equation}
	\label{F0Result}
	\mathcal{F}_{0}(x)= 
	\sum_{L=0}^\infty (-g^2)^L \sum_{l=0}^{L} (-1)^{l} C_{2(L-l)} H_{\ub{2}{l}}(x) \, ,
\end{equation}
where $H_{\emptyset}(x) = 1$ and the constants $C_l$ are proportional to Bernoulli numbers $B_l$,
\begin{equation}
	C_l = \Big| \frac{(2^l-2) \pi^l B_l}{l!} \Big| \, .
\end{equation}
For example, the first few orders of this quantity are given by 
\begin{align}
	\mathcal{F}_{0}(x)= 1 - g^2 \left( \frac{\pi^2}{6} - H_2(x) \right) +
	g^4 \left( \frac{7\pi^4}{360} - \frac{\pi^2}{6}H_2(x) + H_{2,2}(x) \right) \nonumber \\ -
	g^6 \left( \frac{31\pi^6}{15120} - \frac{7\pi^4}{360}H_2(x) + \frac{\pi^2}{6}H_{2,2}(x) - H_{2,2,2}(x) \right) + O(g^8)\,.
\end{align}
We have checked formula~\eqref{F0Result} up to $\mathcal{O}(g^{24})$.

Moving on to the $\alpha\ne 0$ case, it is easy to show that the general recursion formula \eqref{eq:Spq_recursion}, restricted to $N\to \infty$ as in \eqref{eq:EZ_Spq_recursion}, simplifies to
\begin{equation}
	\label{SResult}
	S_{i,j}(\a|x) =
	\frac{1}{\a} \sum_{\mu=1}^{\a-1} S_{i,j-1}(\mu|x) +
	\frac{x^{-\a}}{\a} \sum_{\mu=2}^{\a} x^\mu S_{i-1,j}(\mu|x) + V_{i,j}(\a|x) \, .
\end{equation}
The boundary term 
\begin{align}
	V_{i,j}(\a|x) &\equiv {\cal V}^{1,1}_{\ub{1}{i};\ub{1}{j}}(\a|x) 
\end{align}
similarly reduces to
\begin{align}
	V_{i,j}(\a|x)  &= \frac{1}{\a} \sum_{n=1}^{\infty} \frac{x^n}{n} Z_{\ub{1}{i}}(n-1) Z_{\ub{1}{j}}(n) -
	\frac{x^{1-\a}}{\a} \sum_{n=1}^{\infty} \frac{x^n}{n+1} Z_{\ub{1}{i}}(n-1) Z_{\ub{1}{j}}(n) \nonumber \\
	&\qquad \qquad +\delta_{i,0} \frac{x^{-\a}}{\a} 
	\Big[Z_{\ub{1}{j+1}}(\a|x, \underbrace{1,\ldots,1}_{j} ) - \delta_{j,0}x \Big]\, ,
\end{align}
and can be expressed in terms of HPLs. Finally, the recursion terminates when both $i$ and $j$ are zero, at which point~\eqref{Sk_2} evaluates to
\begin{align}
	S_{0,0}(\a|x) &= 
	\frac{x^{-\a}-1}{\a} \log(1-x) + \frac{x^{-\a}}{\a} Z_1(\a|x) \,.
\end{align}
Due to the fact that $\epsilon$ depends on $g^2$, we need to evaluate $S_{i,j}(\a|x)$ for all $i+j\le L-1$ to compute $\ol$. We have explicitly carried out these sums for all $i+j \leq 11$. These results are easy to verify numerically for various values of $\alpha$ by truncating the sum over $n$ in equation \eqref{Sk}.

\subsection{Resumming \texorpdfstring{$\Omega^{(L)}$}{Omega} through \texorpdfstring{$L=10$}{L=10} loops}

\begingroup
\allowdisplaybreaks

Having evaluated the expansion of $\mathcal{F}_{-i\alpha}(x)$ in terms of $Z$-sums, we now carry out the outermost sum in equation~\eqref{eq:OmegaSum}. Denoting the perturbative expansion of this function as
\be\label{eq:CalFExpansion}
	\mathcal{F}_{-i\alpha}(x) = \sum_{l=0}^{\infty} (-g^2)^l \mathcal{F}^{(l)}_{-i\alpha}(x), \
\ee 
we may exploit the $x\leftrightarrow y$ and $z\leftrightarrow1/z$ symmetry of $\ol$ by adopting the notation
\be{\label{fl}
	f_{\alpha}^{(l)}(x,y) \equiv \frac{1}{1+\delta_{l,L/2}}\mathcal{F}^{(L-l)}_{-i\alpha}(x) \; \mathcal{F}^{(l)}_{-i\alpha}(y)\,,\qquad \alpha\ge 0\,,}
\ee
as well as by introducing the building blocks
\begin{align}\label{eq:om_def}
	\omega^{(L)}(r,x,y) \equiv \sum_{\alpha=1}^{\infty} r^\alpha \sum_{l=0}^{\floor{L/2}} f_\alpha^{(l)}(x,y), \qquad
	\omega^{(L)}_0(x,y) \equiv \sum_{l=0}^{\floor{L/2}} f_0^{(l)}(x,y)\,,
\end{align}
where $\floor{a}$ denotes the integer part of $a$. The sum representation of $\ol$ in~\eqref{eq:OmegaSum} may then be written as
\begin{align}
	\ol(x,y,z) = - \sum_{\alpha=1}^{\infty} (r^\alpha + R^\alpha) \sum_{l=0}^{L} f_\alpha^{(l)}(x,y) - \sum_{l=0}^{L} f_0^{(l)}(x,y) \,+\, \big(x\leftrightarrow y\big)\,,
\end{align}	
where 
\begin{align}
r\equiv - \sqrt{x y z}, \qquad R\equiv - \sqrt{\frac{x y}{z}} \, ,
\end{align}
and finally as
\begin{align} \label{eq:omega_final_sum}
	\ol(x,y,z) = - \omega^{(L)}(r,x,y) - \omega^{(L)}_0(x,y)\,+\, \big(x\leftrightarrow y\big)\,+\, \big(r \leftrightarrow R\big).
\end{align}
Note that the variables $r$ and $R$ used here differ from similarly-named variables used in~\cite{Caron-Huot:2018dsv} by a minus sign. To evaluate the double pentaladder integral in terms of generalized polylogarithms at a given loop order, we thus proceed as follows:
\begin{enumerate}
	\item[(1)] We evaluate the perturbative coefficients $\mathcal{F}^{(l)}_{-i\alpha}(x)$ in terms of $Z$-sums for $l \leq L$.
	\item[(2)] Forming the products $f_{\alpha}^{(l)}(x,y)$, we use stuffle relations to express them as linear combinations of individual $Z$-sums.
	\item[(3)] The results for $f_{\alpha}^{(l)}(x,y)$ can now be substituted into $\omega^{(L)}(r,x,y)$ and $\omega_0^{(L)}(x,y)$, and the remaining infinite sums may be identified as generalized polylogarithms using~\eqref{eq:Z_sums_to_polylogs}.
	\item[(4)] The final result for $\ol$ is then given by~\eqref{eq:omega_final_sum}.
\end{enumerate}
In this manner, we have been able to obtain explicit expressions for the double pentaladder integral up to $L=10$ loops; beyond this loop order, applying the stuffle relations in step (2) becomes computationally onerous. The size of these expressions grows quickly with $L$---at one through ten loops, the output of equation~\eqref{eq:omega_final_sum} involves $\{23, 130, 653, 3205, 15562,$ $ 74717, 352153, 1626600,7372681, 32873641\}$ independent terms. Here we quote the building blocks $\omega^{(L)}$ and $\omega_0^{(L)}$ for one loop,
\begin{align}
	\omega^{(1)}_0(x,y,r) &=
\frac{\pi ^2}{6}-\text{Li}_2(x) \, ,\\
	\omega^{(1)}(x,y,r) &=
-\text{Li}_{1,1}(r,x)+\text{Li}_{1,1}\left(\frac{r}{x},x\right)-\text{Li}_{1,1}(x,r)+\text{Li}_{1,1}(1,r)-\text{Li}_2(r x)+\text{Li}_2(r) \, , \!\!\!
\end{align}
and for two loops,
\begin{align}
	2\omega^{(2)}_0(x,y,r) &=
\text{Li}_{2,2}(x,y)+\text{Li}_{2,2}(y,x)+2 \text{Li}_{2,2}(1,x)+\text{Li}_4(x y) \nonumber\\
&\qquad \qquad -\frac{\pi^2}{2} \text{Li}_2(x)-\frac{\pi^2}{6} \text{Li}_2(y)+\frac{\pi^4}{15}  \, ,
\end{align}
\begin{align}
\!\!\!\!\!\!\! 2\omega^{(2)}(x,y,r) =& \;
2 \text{Li}_{2,2}\left(x^2,\frac{r}{x}\right)+\text{Li}_{1,3}(x,r y)+\text{Li}_{1,3}(y,r x)+\text{Li}_{2,2}(r,x y)-\text{Li}_{2,2}\left(\frac{r}{x},x y\right)\nonumber\\ &-\text{Li}_{2,2}\left(\frac{r}{y},x y\right)+\text{Li}_{2,2}(x y,r)+\text{Li}_{2,2}\left(\frac{r}{x y},x y\right)-\text{Li}_{3,1}\left(\frac{r y}{x},x\right)+\text{Li}_{3,1}(r x,y)\nonumber\\ &-\text{Li}_{3,1}\left(\frac{r x}{y},y\right)+\text{Li}_{3,1}(r y,x)+\text{Li}_{1,1,2}(x,y,r)+\text{Li}_{1,1,2}(y,x,r)+\text{Li}_{1,2,1}(x,r,y)\nonumber\\ &-\text{Li}_{1,2,1}\left(x,\frac{r}{y},y\right)+\text{Li}_{1,2,1}(y,r,x)-\text{Li}_{1,2,1}\left(y,\frac{r}{x},x\right)+\text{Li}_{2,1,1}(r,x,y)+\text{Li}_{2,1,1}(r,y,x)\nonumber\\ &-\text{Li}_{2,1,1}\left(\frac{r}{x},x,y\right)-\text{Li}_{2,1,1}\left(\frac{r}{x},y,x\right)-\text{Li}_{2,1,1}\left(\frac{r}{y},x,y\right)-\text{Li}_{2,1,1}\left(\frac{r}{y},y,x\right)\nonumber\\ &+\text{Li}_{2,1,1}\left(\frac{r}{x y},x,y\right)+\text{Li}_{2,1,1}\left(\frac{r}{x y},y,x\right)-3 \text{Li}_{1,3}(1,r x)+2 \text{Li}_{1,3}\left(\frac{1}{x},r x\right)\nonumber\\ &+3 \text{Li}_{1,3}(x,r)-2 \text{Li}_{2,2}(r,x)-5 \text{Li}_{2,2}(x,r)+\text{Li}_{3,1}(r,x)-\text{Li}_{3,1}\left(\frac{r}{x},x\right)-3 \text{Li}_{1,1,2}(1,x,r)\nonumber\\ &-4 \text{Li}_{1,1,2}\left(1,x,\frac{r}{x}\right)+2 \text{Li}_{1,1,2}\left(\frac{1}{x},x,r\right)-3 \text{Li}_{1,1,2}(x,1,r)+4 \text{Li}_{1,1,2}\left(x,x,\frac{r}{x}\right)\nonumber\\ &-3 \text{Li}_{1,2,1}(1,r,x)-\text{Li}_{1,2,1}\left(1,\frac{r}{x},x\right)+2 \text{Li}_{1,2,1}\left(\frac{1}{x},r,x\right)+2 \text{Li}_{1,2,1}\left(x,\frac{r}{x},x\right)\nonumber\\ &-\text{Li}_{1,3}(1,r y)-\text{Li}_{1,3}(y,r)-\text{Li}_{2,2}(y,r)-\text{Li}_{3,1}(r,y)+\text{Li}_{3,1}\left(\frac{r}{y},y\right)-\text{Li}_{1,1,2}(1,y,r)\nonumber\\ &-\text{Li}_{1,1,2}(y,1,r)-\text{Li}_{1,2,1}(1,r,y)+\text{Li}_{1,2,1}\left(1,\frac{r}{y},y\right)-2 \text{Li}_{1,3}(1,r)+\text{Li}_{2,2}(1,r)\nonumber\\ &+4 \text{Li}_{1,1,2}(1,1,r)+\text{Li}_4(r x y)-\text{Li}_4(r x)-\text{Li}_4(r y)+\frac{\pi^2}{3} \text{Li}_2(r)-\text{Li}_4(r) \, .
\end{align}
Expressions through six loops are provided as ancillary computer files with the \texttt{arXiv} submission of this article. Readers interested in the results for seven, eight, nine, or ten loops may contact the authors. 

Finally, we note that similar sum representations for three additional integrals were derived in~\cite{Caron-Huot:2018dsv}. Our evaluation method works equally well in these cases. However, these integrals are related to the derivatives of $\ol$, so explicit polylogarithmic representations for them can be derived more efficiently by starting from the already-resummed representation of $\ol$.
\endgroup

\subsection{Comparison to Existing Results in the Literature}

Before closing this section, let us compare our results for $\Omega^{(L)}$ with the existing results in the literature. As mentioned above, the authors of~\cite{Caron-Huot:2018dsv} were able to resum $\Omega^{(L)}$ in the $u\to1$ limit and the $w\to 0$ limit. We have compared to these results by taking the appropriate limits of our expressions through four loops, and find complete agreement.

The double pentaladder integral $\Omega^{(L)}$ was also computed in general kinematics via direct integration through four loops in~\cite{Bourjaily:2018aeq}. The comparison of our results with the expressions presented there is a bit subtle, as these representations are manifestly real in different regions. The sum representations of $\Omega^{(L)}$ in~\eqref{eq:omega_final_sum} converges when $x,y, |r| , |R|\le 1$, and is thus manifestly real when $z\sim 1$ and $x, y \ll 1$. This corresponds to the neighborhood of the point $u = v = w = 1$.\footnote{This is easiest to see from equation (A.10) of~\cite{Caron-Huot:2018dsv}.} Conversely, the expressions given in~\cite{Bourjaily:2018aeq} are manifestly convergent in the so-called positive region, which corresponds to positive values of the variables $f_1$, $f_2$, and $f_3$ used in that paper. In particular, the point $u=v=w=1$ corresponds in those variables to the limit $f_1, f_3 \rightarrow -1, f_2 \rightarrow\infty$, well outside of the positive region. 

To compare these representations, we analytically continue the expressions for $\ol$ in terms of $f_1$, $f_2$, and $f_3$ out of the positive region to the neighborhood of the point $u = v = w = 1$. To do so, we note that when $f_1$, $f_2$, and $f_3$ are all small and positive, $u$ $v$, and $w$ are also all positive. Thus, the required analytic continuation path, which changes the signs of $f_1$ and $f_3$, entirely exists within the Euclidean region. This implies that the signs of the cross-ratios $u$, $v$, and $w$ should not change, which turns out to be true only if $f_1$ and $f_3$ are analytically continued in the opposite direction. After sending $f_1 \to f_1 e^{\pm i \pi}$ and $f_3 \to f_3 e^{\mp i \pi}$, we indeed find that the resulting expression is manifestly real near $u = v = w = 1$. This then allows us to confirm that the two representations match in this region.

\section{Application II: Self-energy Diagram} \label{sec:self_energy}

Another natural use of the resummation algorithm presented in section~\ref{subsec:AlgoStatement} is the expansion of one-loop integrals in dimensional regularization. In this section we illustrate how this can be done for families of massive self-energy diagrams with different propagator powers.

\begin{figure}[t]
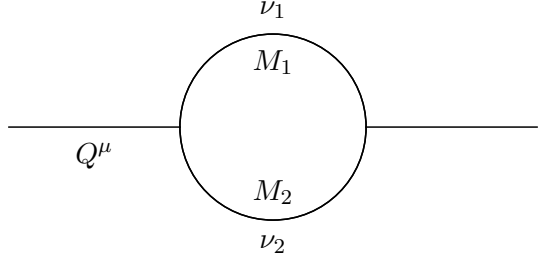

\begin{center}
\begin{fmfchar*}(200,75)
  \fmfleft{ve1}
  \fmfright{ve4}
  \fmf{plain,width=.2mm,label=$Q^\mu$}{ve1,ve2}
  \fmf{plain,width=.2mm}{ve3,ve4}
  \fmf{plain,width=.2mm,right,label=$\nu_1$,tension=.23}{ve3,ve2}
  \fmf{plain,width=.2mm,right,label=$\nu_2$,tension=.23}{ve2,ve3}
  \fmf{plain,width=.2mm,right,label=$M_1$,label.side=left,tension=.23}{ve3,ve2}
  \fmf{plain,width=.2mm,right,label=$M_2$,label.side=left,tension=.23}{ve2,ve3}
\end{fmfchar*}
\end{center}
\caption{Self-energy diagram with masses $M_1$ and $M_2$, and  propagator powers $\nu_1$ and $\nu_2$.}
\label{fig:self_energy}
\end{figure}

We depict a generic massive self-energy diagram with propagator powers $\nu_1$ and $\nu_2$ in figure~\ref{fig:self_energy}. This diagram was evaluated in \cite{Anastasiou:1999ui}, where it was expressed in terms of Appell's $F_4$ function. It was also studied there in various kinematic limits, including the limit of zero external momentum, $Q^2 \to 0$. In that limit, the diagram can be written in terms of Gauss hypergeometric functions. For $M_1>M_2$, it is given by
\begin{align}
	I_2^D(\nu_1,&\nu_2; Q^2 \to 0, M_1^2, M_2^2) =\nonumber\\ 
	&(-1)^{\frac{D}{2}} (-M_1^2)^{\frac{D}{2}-\nu_1-\nu_2} 
	\frac{\Gamma\left(\nu_1+\nu_2-\tfrac{D}{2}\right) \Gamma\left(\tfrac{D}{2}-\nu_2\right)}{\Gamma(\nu_1) \Gamma\left(\tfrac{D}{2}\right)}  
	{}_2F_1\left(\nu_2, \nu_1+\nu_2-\tfrac{D}{2}, 1+\nu_2-\tfrac{D}{2} | x\right) \nonumber \\  &\quad +
	(-1)^{\frac{D}{2}} (-M_1^2)^{-\nu_1} (-M_2^2)^{\frac{D}{2}-\nu_2}
	\frac{\Gamma\left(\nu_2-\tfrac{D}{2}\right)}{\Gamma(\nu_2)}  
	{}_2F_1\left(\nu_1, \tfrac{D}{2}, 1+\tfrac{D}{2}-\nu_2 | x\right)  \, ,
\end{align}
where $\smash{x=\frac{M_2^2}{M_1^2}}$ and $D=4-2\eps$. For $M_1<M_2$ the expression is the same, but with the two masses exchanged.

The function $I_2^D(\nu_1,\nu_2; 0, M_1^2, M_2^2)$ depends on two integers, the powers of the self-energy diagram propagators $\nu_1$ and $\nu_2$. Thus, to make use of the methods of section~\ref{sec:Algorithm} we must constrain $\nu_1$ and $\nu_2$ to depend on only a single symbolic integer. For concreteness, we impose the relation
\begin{align}
\nu_1+\nu_2=4\,, \label{eq:self_energy_restriction}
\end{align}
although we could have chosen any integer on the right hand side.

\begingroup
\allowdisplaybreaks

After imposing the restriction~\eqref{eq:self_energy_restriction}, we expand the $I_2^D(\nu_1,\nu_2; 0, M_1^2, M_2^2)$ around small values of $\epsilon$ using~\eqref{eq:2F1Expansion}. This gives us
\begin{align}
	I_2^D(\nu, 4 - \nu; 0, M_1^2, M_2^2) &=
	(-1)^{-\eps} (-M_1^2)^{-2-\eps} 
	\frac{\Gamma(-\nu +\epsilon +3) \Gamma(\nu -\epsilon -2)}{\Gamma(\nu) \Gamma(4-\nu)\Gamma(2-\eps)}  \, 
	f_{\eps}(4-\nu|x) \\&+
	(-1)^{-\eps} (-M_1^2)^{-\nu} (-M_2^2)^{\nu -\epsilon -2}
	\frac{\Gamma(\nu-\eps-1)\Gamma(-\nu +\epsilon +2)}{\Gamma(\nu)\Gamma(4-\nu)\Gamma(2-\eps)} \,
	f_{-\eps}(\nu|x)\, , \nonumber
\end{align}
where
\begin{equation} \label{eq:self_energy_f}
	f_{\eps}(\nu|x) =	
	\sum_{n=0}^{\infty}x^n (n+1)(n+\nu-1)\sum_{i_1,i_2=0}^{\infty}(-1)^{i_2} \eps^{i_1+i_2}
	Z_{\ub{1}{i_1}}(n+1) S_{\ub{1}{i_2}}(n+\nu-2) \, .
\end{equation}
Using the methods described in sections~\ref{sec:Zsums} and~\ref{sec:Algorithm}, $f_{\eps}(\nu)$ can be converted into $Z$-sums for generic integer values of $\nu$. We have explicitly carried out this calculation through $\mathcal{O}(\eps^6)$ (which took a reasonable time on a laptop). The result through $O(\eps^3)$ are given by
\begin{align}
	f_\eps(\nu|x) =& 
	- \eps^0 g_3 + 
	\eps^1 \Big\{ -g_4 + Z_1(\nu) g_3 -Z_1(\nu|x) g_1 +\log(1-x)\left[g_3-g_1\right] \Big\} \nonumber \\ & +
	\eps^2 \bigg\{
		2-
		g_5  + 
		\li_2(x)[g_3-g_1] + 
		Z_1(\nu) g_4  + 
		Z_1(\nu|x) g_2  \nonumber \\ & \quad+
		\log(1-x) 
		\big[ 
			g_2 +
			g_4 - 
			Z_1(\nu)[g_3+g_1] + 
			Z_1(\nu|\tfrac{1}{x}) g_3 + 
			Z_1(\nu|x) g_1 
		\big] \nonumber \\ & \quad-
		Z_{1,1}(\nu) g_3 -
		Z_{1,1}(\nu|1,x) g_1 + 
		Z_{1,1}(\nu|\tfrac{1}{x},x) g_3 + 
		Z_{1,1}(\nu|x,1) g_1
	\bigg\} \nonumber \\ & +
	\eps^3 \Bigg\{
		\frac{1}{\nu} +
		\li_3(x)\big[g_1-g_3\big] + 
		Z_1(\nu) [g_5-2] -  
		Z_1(\nu|x) g_6+ 
		 \nonumber \\ & \quad+
		\li_2(x)
		\big[
			g_2 + 
			g_4 - 
			Z_1(\nu)[g_1+g_3] + 
			Z_1(\nu|\tfrac{1}{x}) g_3 + 
			Z_1(\nu|x) g_1
		\big] \nonumber \\ & \quad+ 
		Z_{1,1}(\nu|1,x) g_2-
		Z_{1,1}(\nu) g_4+
		Z_{1,1}(\nu|\tfrac{1}{x},x)  g_4  - 
		Z_{1,1}(\nu|x,1) g_2  \\ & \quad+ 
		\log(1-x)
		\bigg[
			g_7 -
			g_6 +
			Z_1(\nu) [ g_2 - g_4]  \nonumber \\ & \quad\quad+
			Z_1(\nu|\tfrac{1}{x})  g_4  -
			Z_1(\nu|x) g_2  +
			Z_{1,1}(\nu) [g_3-g_1]-
			Z_{1,1}(\nu|1,\tfrac{1}{x}) g_3  +
			Z_{1,1}(\nu|1,x) g_1 \nonumber \\ & \quad\quad+
			Z_{1,1}(\nu|\tfrac{1}{x},1) g_3 -
			Z_{1,1}(\nu|\tfrac{1}{x},x) g_3 -
			Z_{1,1}(\nu|x,1) g_1 +
			Z_{1,1}(\nu|x,\tfrac{1}{x}) g_1
		\bigg] \nonumber \\ & \quad+ 
		Z_{1,1,1}(\nu)g_3-
		Z_{1,1,1}(\nu|1,1,x) g_1 - 
		Z_{1,1,1}(\nu|1,\tfrac{1}{x},x) g_3 + 
		Z_{1,1,1}(\nu|1,x,1) g_1 \nonumber \\ & \quad+
		Z_{1,1,1}(\nu|\tfrac{1}{x},1,x) g_3 - 
		Z_{1,1,1}(\nu|\tfrac{1}{x},x,1) g_3 - 
		Z_{1,1,1}(\nu|x,1,1) g_1 +	
		Z_{1,1,1}(\nu|x,\tfrac{1}{x},x) g_1 \nonumber
	\Bigg\} \, ,
\end{align}
where 
\begin{align}
	g_1 &= \frac{x^{2-\nu} (3 + x(\nu-1) - \nu)}{(x-1)^3} \, ,\\
	g_2 &= \frac{x^{-\nu}}{\nu(x-1)^2} \big[ x^2(\nu^2+\nu-1) + x(\nu+2) + \nu -1   \big] \, ,\\
	g_3 &= \frac{x(3-\nu) + \nu - 1}{(x-1)^3} \, ,\\
	g_4 &= -1+\frac{1}{\nu}-\frac{\nu}{(x-1)^2}-\frac{3x}{(x-1)^2} \, ,\\ 
	g_5 &= \frac{3 x}{x-1}\, , \\
	g_6 &= \frac{ x^{-\nu}(1+2x)}{x-1}\, , \\
	g_7 &= \frac{2+x}{x-1} \, .
\end{align}
Unlike the sums contributing to the double pentaladders, $f_\eps(\nu|x)$ is not of uniform transcendental weight, and includes contributions ranging from weight 0 to weight $n$ at $\mathcal{O}(\eps^n)$.

\endgroup

\subsection*{Values for particular \texorpdfstring{$(\nu_1,\nu_2)$}{(nu1,nu2)}}

The class of self-energy diagrams with $\nu_1+\nu_2=4$ includes $(\nu_1,\nu_2)=(1,3)$,  $(\nu_1,\nu_2)=(2,2)$, and $(\nu_1,\nu_2)=(3,1)$. Interestingly, for all values of $\nu$ in this range, contributions with nonzero transcendental weight drop out of $f_\eps(\nu|x)$. In addition, the $\eps$ expansion truncates at low order. Specifically, we find
\begin{align}
f_\eps(1|x)&= -\frac{2x}{(x-1)^3}\,\eps^0 + \frac{1+x}{(x-1)^2}\,\eps^1 - \frac{1}{x-1}\,\eps^2 \, ,\\
f_\eps(2|x)&= -\frac{1+x}{(x-1)^3}\,\eps^0 + \frac{1}{(x-1)^2}\,\eps^1 \, ,\\
f_\eps(3|x)&= -\frac{2}{(x-1)^3}\,\eps^0 \,.
\end{align}

\section{Conclusions}\label{sec:conclusions}

In this paper, we have presented a telescoping relation, enabling an algorithm to convert any sum of the form~\eqref{eq:general_class_sums} into nested sums for symbolic values of $\alpha$, $N$, $x$, $y_i$, and $z_i$. For generic values of $N$, these sums evaluate to cyclotomic harmonic sums, while in the $N \to \infty$ limit they reduce to $Z$-sums. This algorithm allows us, in particular, to resum the expansion coefficients of Gauss hypergeometric functions in cases where this expansion is around integer indices that depend on a symbolic integer parameter $\alpha$. While these hypergeometric expansion coefficients cannot be expressed solely in terms of generalized polylogarithms for generic values of $\alpha$, the additional $Z$-sums that appear reduce to rational functions of their arguments for any specific value of $\alpha$.

We have illustrated how this new resummation technology can be applied to the calculation of Feynman integrals in two examples. In the first, we considered the kinematic expansion of the double pentaladder integrals derived in~\cite{Caron-Huot:2018dsv}, and explicitly evaluated this sum in terms of generalized polylogarithms through ten loops. This represents a significant advance over the previous state of the art, which leveraged the integral representation of these diagrams and became computationally infeasible beyond four loops~\cite{Bourjaily:2018aeq}. In the second example, we showed how this technology can be used to simultaneously expand massive one-loop self-energy diagrams with different propagator powers in dimensional regularization. 

Arguably the biggest shortcoming of our algorithm when applied to the expansion of Gauss hypergeometric functions is the requirement that the symbolic integer appear with the same coefficient in the first and third indices, and be absent from the second index. This ensures that no binomial coefficients occur in the expansion of the hypergeometric function. It is our hope that the general telescoping strategy outlined in section~\ref{sec:algorithm_strategy} can also be leveraged to convert sums that depend on symbolic binomial coefficients into $Z$-sums and polylogarithms, but we leave this to future work. In the meantime, we highlight that the telescopic recursion~\eqref{eq:Spq_recursion} is significantly more general than its instantiation in the Gauss hypergeometric case; we anticipate it will find further applications in perturbative quantum field theory computations.

One direction in which new nested resummation technology would prove fruitful is in the Pentagon Operator Product Expansion (POPE) representation of amplitudes in planar $\mathcal{N}=4$ supersymmetric Yang-Mills~\cite{Basso:2013vsa,Basso:2013aha,Basso:2014koa,Basso:2014nra,Basso:2014hfa,Basso:2015rta,Basso:2015uxa}. The POPE represents amplitudes---or rather, their dual null polygonal Wilson loops---at finite coupling in terms of an expansion in flux tube states propagating across the Wilson loop. This representation can be expanded around small coupling, resulting in infinite sum representations for perturbative amplitudes. Building on earlier work~\cite{Papathanasiou:2013uoa,Papathanasiou:2014yva}, the resummation of these expressions was initiated in~\cite{Drummond:2015jea}, and further continued in~\cite{Cordova:2016woh,Lam:2016rel,Belitsky:2017wdo,Belitsky:2017pbb}. Due to the appearance of binomial coefficients, however, this resummation cannot be carried out systematically at higher orders. We are optimistic that a telescoping strategy will be a viable way forward in these cases.

A great deal is known about the analytic properties of Feynman integrals, and it would be interesting to understand the implications of these properties for sum representations of these integrals. For instance, Feynman integrals are expected to obey the Steinmann relations~\cite{Steinmann,Steinmann2,Cahill:1973qp,Caron-Huot:2016owq}, and the double pentaladder integrals are even known to obey an extended set of Steinmann relations to all orders~\cite{Caron-Huot:2018dsv,Caron-Huot:2019bsq}. However, it is not clear how this property is encoded in the representation of the double pentaladder integrals in~\eqref{eq:OmegaSum}. It would also be interesting to find a sum representation of loop-level amplitudes in planar $\mathcal{N}=4$ supersymmetric Yang-Mills that make the observed positivity properties of these amplitudes manifest~\cite{Arkani-Hamed:2014dca,Dixon:2016apl}, or that make connections with the cluster-algebraic properties of these amplitudes~\cite{ArkaniHamed:2012nw,Golden:2013xva,Golden:2014xqa,Drummond:2017ssj,Drummond:2018dfd,Golden:2018gtk,Golden:2019kks,Drummond:2019cxm,Arkani-Hamed:2019rds,Henke:2019hve}.
 
Finally, it is worth highlighting that, while certain quantities in quantum field theory are believed to be expressible in terms of generalized polylogarithms at low multiplicity or loop order (see for instance~\cite{Beisert:2006ez,CaronHuot:2011ky,Bourjaily:2016evz,DelDuca:2016lad,Henn:2016jdu,Almelid:2017qju,Caron-Huot:2019vjl,Dixon:2016nkn,Drummond:2018caf,Bourjaily:2019iqr,Bourjaily:2019gqu}), Feynman diagrams seem to involve integrals of unbounded algebraic complexity as one goes to higher loop orders~\cite{Brown:2009ta,Bloch:2016izu,Bourjaily:2018ycu,Bourjaily:2018yfy,Bourjaily:2019hmc,mirrors_and_sunsets}. Given that any dimensionally-regularized integral can be considered as a generalized  hypergeometric function~\cite{delaCruz:2019skx}, it would be interesting to see the aforementioned types of functions appearing in their expansion. 
\\\\

\noindent\textbf{Acknowledgements}
We are grateful to Sven Moch for useful discussions. Henrik Munch and Georgios Papathanasiou acknowledge support by the Deutsche Forschungsgemeinschaft under Germany’s Excellence Strategy -EXC 2121 ``Quantum Universe'' - 390833306. Matt von Hippel acknowledges the European Union's Horizon 2020 research and innovation program under grant agreement \mbox{No.\ 793151}. Andrew J.~McLeod acknowledges a Carlsberg Postdoctoral Fellowship (CF18-0641). This work was also supported in part by the Danish National Research Foundation (DNRF91), the research grant 00015369 from Villum Fonden, and a Starting Grant \mbox{(No.\ 757978)} from the European Research Council.

\appendix

\section{Derivation of the General Nested Summation Algorithm}
\label{appendix:general_telescoping_algorithm}

In this appendix we derive the general recursion relation~\eqref{eq:Spq_recursion} that allows one to convert any sum of the form~\eqref{eq:general_sums_class} into a linear combination of cyclotomic harmonic sums. The general strategy we pursue is the same one used in section~\ref{subsec:proof}, where we treated the case of Euler-Zagier sums and infinite $N$. Namely, we first partial fraction the general case into a linear combination of sums in which either $p$ or $q$ is zero,
\begin{align} \label{eq:general_partial_fractioned_Spqyz}
	{\cal S}^{p,q}_{\mathbf{m};\mathbf{r}}(\a,N|x;\mathbf{y};\mathbf{z}) &= 
	\sum_{i=0}^{p-1}  \frac{\binom{q-1+i}{q-1}}{(-1)^i \a^{q+i}} {\cal S}^{p-i,0}_{\mathbf{m};\mathbf{r}}(\a,N|x;\mathbf{y};\mathbf{z}) \nonumber \\
	& \qquad +  \sum_{i=0}^{q-1}  \frac{\binom{p-1+i}{p-1}}{(-1)^p \a^{p+i}} {\cal S}^{0,q-i}_{\mathbf{m};\mathbf{r}}(\a,N|x;\mathbf{y};\mathbf{z}) \, ,
\end{align}
and derive recursions for ${\cal S}^{p,0}_{\mathbf{m};\mathbf{r}}(\a,N|x;\mathbf{y};\mathbf{z})$ and ${\cal S}^{0,q}_{\mathbf{m};\mathbf{r}}(\a,N|x;\mathbf{y};\mathbf{z})$ separately. 

\paragraph{Telescoping \texorpdfstring{${\cal S}^{p,0}_{\mathbf{m};\mathbf{r}}(\a,N|x;\mathbf{y};\mathbf{z})$}{Sp0}}~\\[-10pt] 

\noindent The steps involved in the derivation of the recursion relation for $\smash{{\cal S}^{p,0}_{\mathbf{m};\mathbf{r}}(\a,N|x;\mathbf{y};\mathbf{z})}$ are nearly identical to those used to derive~\eqref{eq:Sp0_recursion}. In this case, we find
\begin{align}
	{\cal S}^{p,0}_{\mathbf{m};\mathbf{r}}(\a+1,N|x;\mathbf{y};\mathbf{z})
	&= 	{\cal S}^{p,0}_{\mathbf{m};\mathbf{r}}(\a,N|x;\mathbf{y};\mathbf{z}) + z_1^\alpha {\cal S}^{p,r_1}_{\mathbf{m};\mathbf{r'}}(\a,N|x z_1 ,\mathbf{y},\mathbf{z'}) \, ,\label{eq:Sp0_general_alphaPlusOne}
\end{align}
which implies that $P$ is still zero in the ansatz~\eqref{GeneralTele}. We thus have
\begin{align}	
	\label{eq:Sp0_general_recursion}
	 {\cal S}^{p,0}_{\mathbf{m};\mathbf{r}}(\a,N|x;\mathbf{y};\mathbf{z}) =
	z_1^\alpha \sum_{\mu=1}^{\a-1} {\cal S}^{p,r_1}_{\mathbf{m};\mathbf{r'}}(\mu,N|x z_1;\mathbf{y};\mathbf{z'}) + {\cal S}^{p,0}_{\mathbf{m};\mathbf{r}}(1,N|x;\mathbf{y};\mathbf{z}) \, .
\end{align}
It is easy to check that this reproduces~\eqref{eq:Sp0_recursion} when $z_1 = 1$ and $N \to \infty$, and that it gets the correct answer when $|\mathbf{r}|=0$.
 
\paragraph{Telescoping \texorpdfstring{${\cal S}^{0,q}_{\mathbf{m};\mathbf{r}}(\a,N|x;\mathbf{y};\mathbf{z})$}{S0q}}~\\[-10pt] 

\noindent To find the recursion relation for ${\cal S}^{0,q}_{\mathbf{m};\mathbf{r}}(\a,N|x;\mathbf{y};\mathbf{z})$, we first consider
\begin{align} \label{eq:S0q_general_shifted}
	{\cal S}^{0,q}_{\mathbf{m};\mathbf{r}}(\a+1,N|x;\mathbf{y};\mathbf{z}) &= 
	\sum_{n=1}^{N+1} \frac{x^{n-1}}{(n+\a)^q} Z_{\mathbf{m}}(n{-}2|\mathbf{y}) Z_{\mathbf{r}}(n{+}\a{-}1|\mathbf{z})\, ,
\end{align}
where we have shifted the summation index  $n \to n{-}1$ relative to the definition~\eqref{eq:general_sums_class}, and then used the fact that the shifted summand is zero when $n=1$ to change the lower summation bound back to $1$. To increment the upper summation bound of $Z_{\mathbf{m}}(n{-}2|\mathbf{y})$, we substitute~\eqref{eq:Z_arg_shift} into~\eqref{eq:S0q_general_shifted} to get
\begin{align}
{\cal S}^{0,q}_{\mathbf{m};\mathbf{r}}(\a+1,N|x;\mathbf{y};\mathbf{z}) 	
        &= \frac{1}{x} {\cal S}^{0,q}_{\mathbf{m};\mathbf{r}}(\a,N+1|x,\mathbf{y},\mathbf{z}) - {\cal S}^{m_1,q}_{\mathbf{m'};\mathbf{r}}(\a+1,N|x y_1,\mathbf{y'},\mathbf{z}) \nonumber \\
	&\qquad \qquad - \frac{\delta_{|\mathbf{m}|,0} }{(\a+1)^q} Z_{\mathbf{r}}(\a|\mathbf{z})  \, ,
\end{align}
where we have shifted the index $n \to n{+}1$ in the second term to put it in this form.

Comparing to~\eqref{eq:delta_S}, we see that $P={-}1$. After separating out the $n=N+1$ contribution from ${\cal S}^{0,q}_{\mathbf{m};\mathbf{r}}(\a,N+1|x,\mathbf{y},\mathbf{z})$, we thus have
\begin{align}
	\Delta {\cal S}^{0,q}_{\mathbf{m};\mathbf{r}}(\a,N|x;\mathbf{y};\mathbf{z}) &\equiv {\cal S}^{0,q}_{\mathbf{m};\mathbf{r}}(\a+1,N|x;\mathbf{y};\mathbf{z}) - \frac{1}{x} {\cal S}^{0,q}_{\mathbf{m};\mathbf{r}}(\a,N|x;\mathbf{y};\mathbf{z}) \\
	&= - {\cal S}^{m_1,q}_{\mathbf{m'};\mathbf{r}}(\a+1,N|x y_1;\mathbf{y'};\mathbf{z})  - \frac{\delta_{|\mathbf{m}|,0}}{(\a+1)^q} Z_{\mathbf{r}}(\a|\mathbf{z}) \nonumber \\
	&\qquad \qquad+ \frac{x^N}{(N+\a+1)^q} Z_{\mathbf{m}}(N|\mathbf{y}) Z_{\mathbf{r}}(N{+}\a|\mathbf{z}) \, .
\end{align}
Plugging this into~\eqref{GeneralTele}, we find the relation  
\begin{align}
	\label{eq:S0q_general_recursion}
	{\cal S}^{0,q}_{\mathbf{m};\mathbf{r}}(\a,N|x;\mathbf{y};\mathbf{z}) &= 
	- x^{-\a} \sum_{\mu=2}^{\a} x^{\mu} {\cal S}^{m_1,q}_{\mathbf{m'};\mathbf{r}}(\mu,N|xy_1;\mathbf{y'};\mathbf{z})+ x^{1-\a}  {\cal S}^{0,q}_{\mathbf{m};\mathbf{r}}(1,N|x;\mathbf{y};\mathbf{z})   \nonumber \\	
	&\qquad \qquad - x^{-\a} \delta_{|\mathbf{m}|,0} \Big( Z_{q,\mathbf{r}}(\alpha|x,\mathbf{z})  - x \delta_{|\mathbf{r}|,0} \Big) \\
	&\qquad \qquad + x^{-\a} Z_{\mathbf{m}}(N|\mathbf{y}) \Big(Z_{q,\mathbf{r}}(N{+}\a|x,\mathbf{z}) - Z_{q,\mathbf{r}}(N+1|x,\mathbf{z}) \Big) \, , \nonumber
\end{align}
where we have already converted most of the sums over $\mu$ into $Z$-sums, and have shifted $\mu \to \mu{+}1$ in the remaining sum to simplify the summand. Notice that the third line encodes a contribution that was absent in~\eqref{eq:S0q_recursion}, which drops out when $N \to \infty$.

\paragraph{A Closed Recursion for \texorpdfstring{${\cal S}^{p,q}_{\mathbf{m};\mathbf{r}}(\a,N|x) $}{Spq}}~\\[-10pt] 

\noindent Substituting equations \eqref{eq:Sp0_general_recursion} and \eqref{eq:S0q_general_recursion} into \eqref{eq:general_partial_fractioned_Spqyz}, we obtain the recursion relation~\eqref{eq:Spq_recursion} presented in section~\ref{subsec:AlgoStatement}. The sums over $i$ can be performed explicitly for fixed integer values of $p$ and $q$, while the remaining sums can be converted into nested sums. As our most general examples involve two symbolic integer parameters $\a$ and $N$, this requires invoking the class of cyclotomic harmonic sums. We work through an example involving cyclotomic harmonic sums in appendix~\ref{appendix:cyclotomic_example}. In general, this procedure gives rise to a linear combination of generalized polylogarithms, $Z$-sums, and cyclotomic harmonic sums with coefficients that depend on the symbolic parameters $\alpha$, $N$, $x$, $\mathbf{y}$, and $\mathbf{z}$.

\section{An Example Involving Cyclotomic Harmonic Sums}
\label{appendix:cyclotomic_example}

In this appendix we illustrate how cyclotomic harmonic sums \cite{Ablinger:2011te} appear when the upper summation bound in~\eqref{eq:general_sums_class} is left generic by working though the example of ${\cal S}^{1,1}_{\emptyset;1}(\a,N|x;\emptyset;z_1)$. To carry out this sum, we need only apply the recursion identity once. This give us
\begin{align}
{\cal S}^{1,1}_{\emptyset;1}(\a,N|x;\emptyset;z_1) &= \frac{1}{\a}  \sum_{\mu=1}^{\a-1}  \, z_1^\mu {\cal S}^{1,1}_{\emptyset;\emptyset}(\mu,N|x z_1;\emptyset;\emptyset) \label{eq:appendix_example_eq_2}  \\
	&\qquad \qquad + {\cal V}^{1,1}_{\emptyset;1}(\a,N|x;\emptyset;z_1) \, .  \nonumber
\end{align}
Plugging these values into~\eqref{eq:SpqTermination}, we can express the terminal sum as 
\begin{align}
	{\cal S}^{1,1}_{\emptyset;\emptyset}(\mu,N|x  z_1;\emptyset;\emptyset) &=
	\frac{1}{\mu} Z_{1}(N|x  z_1) -
	\sum_{k=1}^N \frac{(x z_1)^{k}}{ \mu (\mu+k)} \, , 
\end{align}
where we have used~\eqref{ZIdentity} to convert the difference of $Z$-sums in the second term into the sum over $k$. Plugging this back into~\eqref{eq:appendix_example_eq_2} and evaluating the sum over $\mu$, one finds
\begin{align}
{\cal S}^{1,1}_{\emptyset;1}(\a,N|x;\emptyset;z_1) &= \frac{1}{\a} \sum_{k=1}^N \frac{x^{k}}{k} \left( Z_1(\a{+}k{-}1 | z_1) - Z_1(k|z_1) - z_1^k Z_1(\a{-}1|z_1)  \right)  \label{eq:appendix_example_eq_3} \\
&\qquad + \frac{1}{\a} Z_{1}(\a{-}1|z_1) Z_{1}(N|x z_1) + {\cal V}^{1,1}_{\emptyset;1}(\a,N|x;\emptyset;z_1) \, .\nonumber 
\end{align}
This sum over $k$ cannot be carried out in terms of $Z$-sums. Rather, we are required to make use of cyclotomic harmonic sums, defined by
\begin{align}
S_{\{a_1,b_1,c_1\},\dots,\{a_d,b_d,c_d\}}(\mathbf{x}; N) = \sum_{n=1}^N \frac{x_1^{n}}{(a_1 n + b_1)^{c_1}} S_{\{a_2,b_2,c_2\},\dots,\{a_d,b_d,c_d\}}(\mathbf{x'}; n) \, ,
\end{align}
where $\mathbf{x} = x_1,\dots,x_d$ is a multi-index of depth $d$, and $S(N) = 1$. As with the $Z$-sums, cyclotomic harmonic sums satisfy a large number of identities, such as stuffle and synchronization identities; they can also be given an iterated integral representation. 

After carrying out the conversion $Z_1(\a {+} k{-}1|z_1) = Z_1(\a{-}1|z_1) + z^{\a-1}S_{\{1,\a{-}1,1\}}(z_1;k)$, the sum over $k$ in~\eqref{eq:appendix_example_eq_3} can be evaluated to give
\begin{align}
{\cal S}^{1,1}_{\emptyset;1}(\a,N|x ;\emptyset;z_1) &= \frac{z^{\a-1}}{\a} S_{\{1,0,1\},\{1,\a{-}1,1\}}(x ,z_1; N) + \frac{1}{\a} Z_1(N|x ) Z_1(\a{-}1| z_1) \label{eq:appendix_example_eq_4}  \\
&\qquad - \frac{1}{\a} Z_{1,1}(N|x , z_1) - \frac{1}{\a} Z_2(N|x  z_1)  + {\cal V}^{1,1}_{\emptyset;1}(\a,N|x;\emptyset;z_1) \, ,\nonumber 
\end{align}
while the boundary term can be evaluated using the same methods as usual:
\begin{align}
{\cal V}^{1,1}_{\emptyset;1}(\a,N|x;\emptyset;z_1) &= \frac{1}{\a} Z_{1,1}(N| x, z_1) + \frac{1}{\a} Z_2(N| x  z_1)  \\
&\qquad + \frac{x^{-\a}}{\a} Z_{1,1}(\a| x, z_1)  - \frac{x^{-\a}}{\a} Z_{1,1}(N{+}\a| x, z_1) \nonumber \, .
\end{align}
Putting this all together, we find
\begin{align}
{\cal S}^{1,1}_{\emptyset;1}(\a,N|x ;\emptyset;z_1) &= \frac{z^{\a-1}}{\a} S_{\{1,0,1\},\{1,\a{-}1,1\}}(x ,z_1; N) + \frac{1}{\a} Z_1(N|x ) Z_1(\a{-}1| z_1) \label{eq:appendix_example_eq_5}  \\
&\qquad + \frac{x^{-\a}}{\a} Z_{1,1}(\a| x, z_1)  - \frac{x^{-\a}}{\a} Z_{1,1}(N{+}\a| x, z_1) \nonumber \, .
\end{align}
It is easy to check that this expression numerically reproduces~\eqref{eq:example_eq_3} (with $y_1$ set to $1$) for large values of $N$.

\bibliographystyle{JHEP}

\bibliography{omega_refs}

\end{fmffile}
\end{document}